\documentclass[smallextended]{svjour3}       
\smartqed  
\usepackage{graphicx}

\usepackage[latin1]{inputenc}
 
\usepackage{url}

\usepackage{setspace}

\usepackage{fancybox}

\usepackage{amssymb}

\usepackage[round]{natbib}
\usepackage{makerobust}
\DeclareRobustCommand{\cite}[2][]{\citep[#1]{#2}}
\DeclareRobustCommand{\citeN}[2][]{\citet[#1]{#2}}

\usepackage{amssymb}
\usepackage{mathptmx}      
\journalname{   }

\begin{document}

\title{An experimental comparison of label selection methods for hierarchical document clusters
}


\author{Maria Fernanda Moura \and
        Fabiano Fernandes dos Santos \and
				Solange Oliveira Rezende
}


\institute{M.F. Moura \at
           Embrapa Agriculture Informatics \\
           PO Box 6041, Campinas, SP, Brazil\\
           Tel.: +55-19-9733-1231\\
           Fax: +55-19-3211-5754\\
           \email{fernanda@cnptia.embrapa.br}          
           \and
           F.F. dos Santos and S.O. Rezende\at
           São Paulo University\\
           Computer Science Department\\
           PO Box 668, São Carlos, SP, Brazil\\
           Tel.: +55-16-3373-9646\\
           Fax: +55-16-3271-2238\\ 
           \email{{fabianof,solange}@icmc.usp.br}
}

\date{Received: date / Accepted: date}

\maketitle

\begin{abstract}
The focus of this paper is on the evaluation of sixteen labeling methods for hierarchical document clusters over five datasets. All of the methods are independent from clustering algorithms, applied subsequently to the dendrogram construction and based on probabilistic dependence relations among labels and clusters. To reach a fair comparison as well as a standard benchmark, we rewrote and presented the labeling methods in a similar notation. The experimental results were analyzed through a proposed evaluation methodology based on: (i) data standardization before applying the cluster labeling methods and over the labeling results; (ii) a particular information retrieval process, using the obtained labels and their hierarchical relations to construct the search queries; (iii) evaluation of the retrieval process through precision, recall and F measure; (iv) variance analysis  of the retrieval results to better understanding the differences among the labeling methods; and, (v) the emulation of a human judgment through the analysis of a topic observed coherence measure - normalized Pointwise Mutual Information (PMI). Applying the methodology, we are able to highlight the advantages of certain methods: to capture specific information; for a better document hierarchy comprehension at different levels of granularity; and, to capture the most coherent labels through the label selections. Finally, the experimental results demonstrated that the label selection methods which hardly consider hierarchical relations had the best results. 
\end{abstract}

\section{Introduction}

Topic hierarchies are  helpful in  organizing and browsing documents,  aiding in the exploration of similar document groups or even in the analysis of topic tendencies within a large document collection \cite{Moura:2008,IencoMeo:2008,Escudeiro:2006,Muhr:2010,Aggarwal:2012}. A document topic hierarchy can be provided by human beings based on their domain knowledge as in hierarchical organizations like \textit{Yahoo!} directories or Open Directory Project\footnote{DMOZ - \url{http://www.dmoz.org/.}} \cite{TreeCal:2006,Tang:2008} or it can be automatically learned from data. In both cases, text documents are organized into topics and subtopics, providing an intuitive way for users to explore textual data at different levels of granularity \cite{Marcacini:2012b}. Thus, a relevant problem is how to automatically  construct a useful topic hierarchy dealing with the flood of textual data.

The hierarchical structure can be automatically obtained through a hierarchical clustering algorithm, since these algorithms have been studied extensively in the clustering literature \cite{Aggarwal:2012}. Besides, there are some optimized algorithms for clustering as well as incremental and parallel solutions \cite{aggarwal2013data,Cai:2013,fahad2014survey}. However, few document clustering methods deal with the automatic hierarchical labeling process as well as the hierarchical relations among labels \cite{mao2012automatic,Muhr:2010,MouraRezende2010}. So, it is possible to produce efficient and effective hierarchical document clusters but, in order to achieve a useful topic hierarchy, we need to select significant labels. The automatic  labeling of the obtained hierarchies consists in selecting a set of key terms to each document group, in order to better understand or browse them. In this way, each node is labeled by a list of key terms and each  internal node is defined by its vertical path (i.e., ancestor and child nodes) and its horizontal list of key terms (or descriptors). 

The simplest labeling method selects the most frequent words present in the documents of each cluster. This method reveals the topic at a higher level, but can fail to depict specific details of the cluster \cite{PopeUngar:2000}. 
In a hierarchical document grouping, not only do we need to distinguish an internal node in the tree from its siblings, as in a flat cluster, but also from its parent and its children \cite{manning}. In many approaches, the labels have to be built using all statistically significant terms in the documents of the group. Thus, it is important to evaluate the behavior of different cluster labeling methods along the hierarchy, for different document collections and for different languages, in order to better capture the meaning of each group. Moreover, there are no standard procedures to evaluate the obtained labels, although some attempts have been made \cite{lawrie,mao2012automatic,Kashyap:2005}. Besides, the evaluation is difficult even when a group of volunteers is willing to participate, because it also depends on the task for which the hierarchy is designed \cite{Bast:2005}.

In this paper, we focused on the evaluation of hierarchical cluster labeling methods considering the non existence of a gold standard - e.g. a correctly labeled taxonomy, or a subjective analysis from a domain specialist. Firstly, we supposed that the clustering process evaluation is a resolved problem.  Consequently, we selected only cluster labeling methods wich can be applied subsequently to the hierarchical document cluster process. Highlighting that these methods can be used without any external knowledge or user intervention.  Thus, we proposed and used an experimental methodology to compare the results of some cluster labeling methods considering the number of labels, hierarchical relations among them and the simulation of a human judgment. 

This study identified sixteen methods that should be used by the research community in the future as a benchmark against which new methods should compete.  The evaluated label selection methods can be found in four articles \cite{PopeUngar:2000, Lamirel:2008, Muhr:2010, MouraRezende2010}. In order to reach fair comparisons among benchmarking methods, we re-specified and re-implemented the methods in the same pattern. The obtained specifications are presented in the second section of this paper. 

The evaluation methodology is based on: (i) data standardization before applying the cluster labeling methods and over the labeling results; (ii) a particular information retrieval process, using the obtained labels and their hierarchical relations to construct the search queries; (iii) evaluation of the retrieval process through precision, recall and F measure; (iv) variance analysis  of the retrieval results to better understanding the differences among the labeling methods; and, (v) the calculation and analysis of a coherence measure which is able to simulate human judgment \cite{lau2014machine}. With this experimental methodology we expect to achieve a fair comparison among different cluster labeling methods: to capture specific information; for a better document hierarchy comprehension at different levels of granularity; and, to capture the most coherent labels through the label selections. The methodology is presented in the third section of this paper. 

We experimentally evaluated the sixteen methods for cluster label selection over five datasets. The experimental results show the differences among the methods through the proposed comparison methodology, as presented in the fourth section of this paper. In the fifth section, the results and a brief discussion on future works are summed up.
 
\section{Benchmark's label selection methods}
\label{sec:back}

Before label selection, in order to reach a satisfactory hierarchical clustering of documents, it is necessary to choose and evaluate (i) a model to represent the documents and the selected terms and (ii) a hierarchical clustering process. In the case of (i), the representation model corresponds to the identification of the most discriminative terms in the text collection and the indexing of the text collection, based on those terms. In text mining and information retrieval areas, this process is also named preprocessing and has received a lot of attention in literature as well as in the development of efficient software tools. Regarding (ii), the selection of a hierarchical clustering process and the obtained grouping evaluation are exhaustive tasks that must be performed carefully. As it was mentioned before, a lot of effort has been put on efficient document cluster development, which does not embed or reach the cluster labeling \cite{Feldman:2007}. Therefore, this paper focuses on cluster labeling, assuming that the steps (i) and (ii) were well resolved.  

Firstly, in this section, we explain the label selection task for a hierarchical document cluster, when considering this task independent from the applied cluster algorithm and the existence of previous knowledge. 
To select the methods for this benchmark, we focused on label selection methods which are independent from cluster algorithm and from domain knowledge - such as vocabulary, thesaurus, ontology, etc. 
Then, we describe the label selection methods for hierarchical document clustering proposed by \citeN{Lamirel:2008}, \citeN{PopeUngar:2000}, \citeN{MouraRezende2010} and \citeN{Muhr:2010}, because they satisfy these constraints. Finally, we also decided to include the Most Frequent method in this study because it is widely used. In order to compartmentalize the discussions, we divided the benchmarking methods into methods based on frequency and methods based on probabilistic or statiscal models.  

\subsection{The task of label selection for hierarchical document clusters}

\begin{figure}[!h]
\centering
\includegraphics[width=1\textwidth]{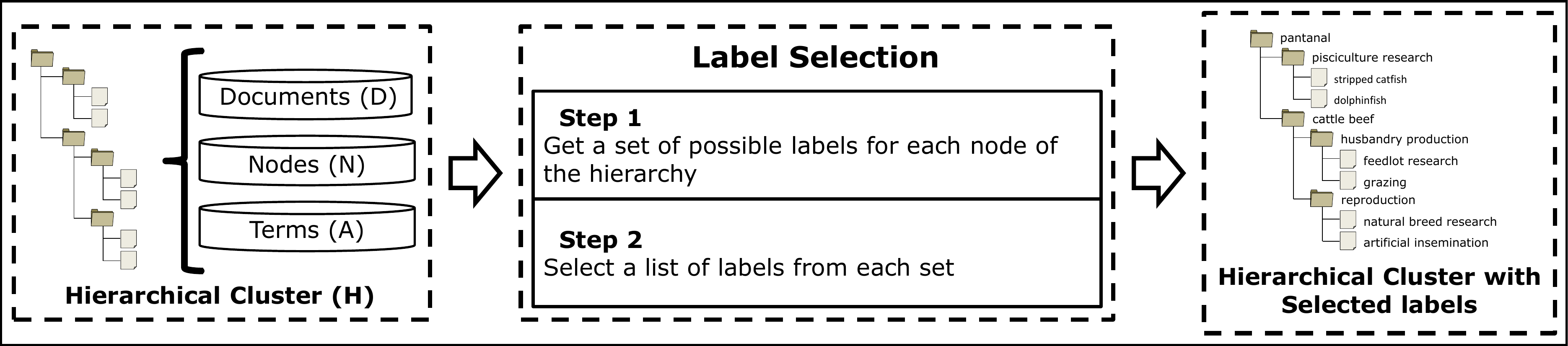}
\caption{Label selection process for hierarchical document cluster \citep{Santos:2012}.}
\label{fig:passos-selecao}
\end{figure}

Generally there are two main steps to the label selection methods for hierarchical document clusters when the labeling method is independent of the cluster algorithm, as observed in Figure \ref{fig:passos-selecao} \cite{Santos:2012}. The input is a hierarchical cluster $H$ built from a set of documents $D$. $A$ is a set of attributes (or terms) previously selected in the preprocessing step and $N$ is the set of nodes in $H$. Initially, the method selects a set of possible labels for each node in $N$ (Step 1). Then some of the possible labels are chosen and the set of labels are built for each node (Step 2).

\begin{figure}[!h]
 	\centering
	\includegraphics[width=0.7\textwidth]{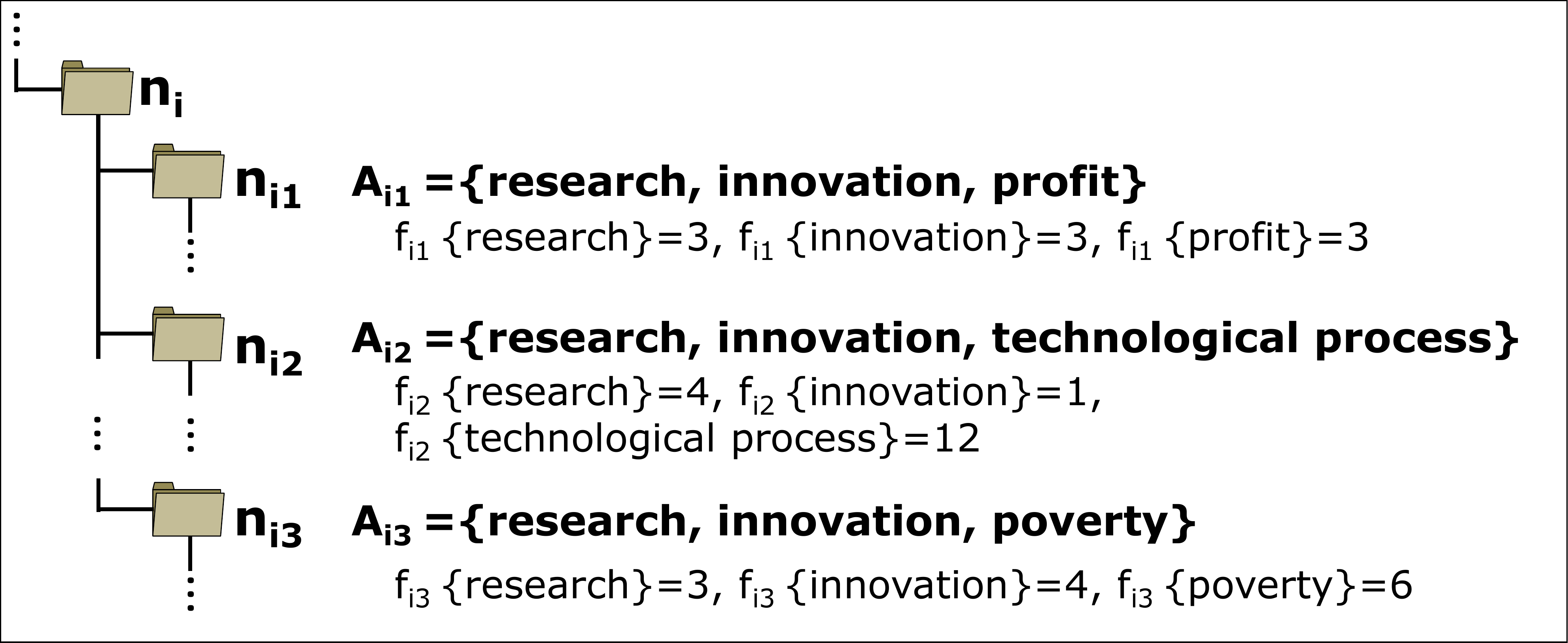}
	\caption{An example of a generic $n_i$ node in a document hierarchy with its $c$ children. Each $n_i$'s children had been already labeled.}
	\label{fig:ExemploTabela3}
\end{figure}

A cluster labeling task independent of the applied clustering method can be viewed as an attribute selection problem. To illustrate this idea and define some of the necessary concepts, Figure \ref{fig:ExemploTabela3} presents the abstraction of a generic node ($n_i$) in a hierarchical structure. The hierarchical structure is supposed to be obtained from a clustering method applied on a document collection, $D=\left\{d_1,...,d_x\right\}$, with $x$ documents, characterized by a set of attributes, $A = \left\{a_1,..a_m\right\}$, with $m$ attributes. Therefore, $N = \left\{n_1,..n_o\right\}$ is the set of nodes and each node $n_i$, $i=1,...,o$, in the hierarchy corresponds to a cluster (or group) with $c_i$ children and is formed by a sub collection from $D$. Thus, $f_i(a_k)$ is defined as the cumulated frequency of the $k^{th}$ attribute from $A$ in $n_i$. In other words, $f_i(a_k)$ corresponds to the cumulated frequency of $a_k$ in the subset of documents in $n_i$. In order to decide which attributes are better to discriminate the $i^{th}$ group, the contingency table illustrated as Table \ref{ContTable} is constructed for each $a_k$, $k=1,..m$ in $n_i$. 

\begin{table}[!h] \centering 
 \caption{Contingency table for $f_i(a_k)$ in $n_i$ child nodes.}
 \label{ContTable} 
\small
 \begin{tabular}{|c||c|c||c|}
  		\hline
			$child$  &     $a_k$      &   $!a_k$                             & total  \\   \hline
      $n_{i1}$ &  $f_{i1}(a_k)$ & $\sum^{m}_{t=1,t \neq k}f_{i1}(a_t)$ & 
                                                    $\sum^{m}_{t=1}f_{i1}(a_t)$ \\  \hline
       $...$   &  $...$         &  $...$                               &  $...$  \\  \hline
      $n_{ic}$ &  $f_{ic}(a_k)$ & $\sum^{m}_{t=1,t \neq k}f_{ic}(a_t)$ & 
                                                    $\sum^{m}_{t=1}f_{ic}(a_t)$ \\ \hline \hline
      $total$    &  $f_i(a_k)$    & $\sum^{m}_{t=1,t \neq k}f_{i}(a_t) $ & 
                                                       $\sum^{m}_{t=1}f_{i}(a_t) $ \\ \hline
  \end{tabular}
\end{table}

To have a better picture of Table \ref{ContTable} values, consider the example in Figure \ref{fig:ExemploTabela3}. In that figure, each child ($n_{i1},n_{i2},n_{i3}$) of the $n_i$ node has a set of preselected labels ($A_{i1}$,$A_{i2}$,$A_{i3}$). Lets consider the $n_{i1}$ label, $A_{i1}=\left\{research,innovation,profit\right\}$. Suppose the $research$ attribute occurs $3$ times in the documents of the $n_{i1}$ group, then the $f_{i1}(research)=3$.  Thus, each attribute selected as a label to each node has a cumulated frequency obtained from the sum of their presence in all documents which are grouped in the node. For example, for the $research$ attribute in the $n_i$ node in Figure \ref{fig:ExemploTabela3}, the contingency table is illustrated in  Table  \ref{researchTable}. The cumulated frequency for $!research$ is the sum of the cumulated frequency for each attribute different from $research$ in the same node, that is, $f_{i1}(!research)=f_{i1}(innovation)+f_{i1}(profit)$ or $f_{i1}(!research)=3+3=6$. 

The attribute selection for each $n_i$ label is based on the frequencies for each attribute observed in a table like Table \ref{ContTable}, considering each document as a bag of words. One way to select labels is through a supervised feature selection considering each group, $n_{ij}$, $i=1,...,o$ and $j=1,...,c_i$, as a class. In this case, mutual information, information gain, $\chi^2$ or other supervised measures can be used to select the attributes. For an unsupervised selection, we can use  \textit{tf-idf (term frequency - inverse document frequency)} or JSD (Jensen-Shannon Divergence) rankings or some latent factors derived from a variance decomposition. 

\begin{table} \centering 
 \caption{Example of the contingency table for $f_i(research)$ distribution among the $n_i$ children nodes from Figure \ref{fig:ExemploTabela3}.}
   \label{researchTable}
\small
   \begin{tabular}{|c||c|c||c|}
  		\hline
			$n_i\ child$  &     $research$      &   $!research$        & total  \\   \hline
            $n_{i1}$      &        $3$          &   $12$               &  $15$  \\  \hline
            $n_{i2}$      &        $4$          &   $13$               &  $17$  \\  \hline
            $n_{i3}$      &        $3$          &   $10$               &  $13$   \\ \hline \hline
            $total$       &       $10$          &   $35$               &  $45$   \\ \hline
  \end{tabular}
\end{table}

In this paper, we are always referring to a generic node representation as in the Figure \ref{fig:ExemploTabela3} and to a generic attribute frequency distribution, in this generic node, as in Table \ref{ContTable}. Finally, in this section, we present all the label selection methods in the same pattern. The basic notation is:

\begin{center}
	$L(n_i) = \{ best_p(A_i), p=1,...,P \}$, where 
\end{center}
%

\begin{center}
	$A_i = \{ \forall a_k \in A  \mid \left( RankingFunction_i(a_k) > 0 \right), k=1,...,m \}$, and
\end{center}
%

\begin{center}
	$RankingFunction_i(a_k)=\ label\ selection\ function\ of\ the\ method$ 
\end{center}

\noindent $A_i\in A$ is a subset of attributes which was selected as a possible label to a $n_i$ node according to a ranking function ($RankingFunction_i$). Finally, $L(n_i)$ is the set of the best $P$ ranked attributes from $A_i$ and $L(n_i)$ is the selected label for the $n_i$ node. $P$ is a predefined natural number and $P\geq1$. Moreover, for the selection functions which do not result in a ranking, the absolute-frequency value of the selected $a_k$ in the $i^{th}$ node was considered as $a_k$ rank.

\subsection{Me\-thods based on frequency}
\label{sec:methods-frequency}

A frequency is associated for each dimension of a document representation. So, the simplest labeling method consists of sorting the terms according to the cumulated frequency of this term in the documents of the $n_i$ node \cite{Miiller,Chuang}. 
%
%
This method is generally called \textbf{Most Frequent} and defined as:

\begin{center}
	
	$L(n_i) = \{ best_p(A_i), p=1,...,P \}$, where
	
\end{center}
%

\begin{center}
	$A_i = \{ \forall a_k \in A \mid \left( MostFrequent_i(a_k) \right) > 0 , k=1,..,m \}$, and
\end{center}
%

\begin{center}
	$MostFrequent_i(a_k) =  \displaystyle\sum\limits_{j=1}^{c} f_{ij}(a_k)$
\end{center}

\citeN{Muhr:2010} explored the structural relationships in hierarchical clusters in order to obtain good labels. They proposed some extensions to standard labeling approaches based on term weights in each node. The standard labeling approach was referred to by \citeN{Muhr:2010} as Maximum Term Weight Labeling (MTWL). The MTWL approach selects the $P$ terms with the highest frequency. \citeN{Muhr:2010} referred to the \textbf{Most Frequent} method as \textbf{MTWL$_{raw}$}. So, to use a single representation, we will refer to the \textbf{Most Frequent} method as \textbf{MTWL$_{raw}$}:

\begin{center}
	$MTWL_{raw_{i}}(a_k) = MostFrequent_i(a_k)$
\end{center}

First, \citeN{Muhr:2010} proposed the Global Inverse Document Frequency (\textit{idf}$_{global}$) and Local Inverse Document Frequency (\textit{idf}$_{local}$) based on traditional Inverse Document Frequency weight. For each node $n_i$, the \textit{idf}$_{global}$ of the $k^{th}$ term is defined as follows:

\begin{center}
	
	$idf_{global}(a_k) = \log \left( \frac{ \left| D \right| }{ \#(a_k , D) } \right)$
	
\end{center}

\noindent where $\left| D \right|$ is the cardinality of $D$ and $\#(a_k , D)$ is the number of documents in $D$ containing term $a_k$ (also named in literature as Document Frequency - df). \citeN{Muhr:2010} claim that \textit{idf}$_{global}$ penalizes terms which are over-represented in the whole collection. However, terms over-represented in a particular cluster sub-tree only will be likely selected for all siblings in the cluster hierarchy. Term distributions among siblings have to be taken into account to avoid siblings getting similar labels. For this reason, the \textit{idf}$_{local}$ weight was proposed by \citeN{Muhr:2010}. The \textit{idf}$_{local}$ for the $k^{th}$ term in cluster $n_{ij}$ with parent cluster $n_i$ is calculated as:

\begin{center}
	
	$idf_{local_{ij}}(a_k) = \log \left( \frac{ \left| D_{n_{i}} \right| }{ \#(a_k , D_{n_{i}}) } \right)$
	
\end{center}

\noindent where $\left| D_{n_{i}} \right|$ is the number of documents in the parent cluster $n_i$ of cluster $n_{ij}$ and $\#(a_k , D_{n_i})$ is the number of documents in the parent cluster $n_i$ containing term $a_k$. These two weight schemes are combined to select the top $P$ terms as labels of cluster $n_{ij}$ and this scheme is nominated by \citeN{Muhr:2010} as MTWL$_{idf}$. Formally, the \textbf{MTWL$_{idf}$} labeling function is defined as:

\begin{center}
	$L(n_{ij}) = \{ best_p(A_{ij}), p=1,...,P \}$, where
	
\end{center}
%

\begin{center}
	
	$A_{ij} = \{ \forall a_k \in A \mid MTWL_{idf_{ij}}(a_k) > 0 , k=1,..,m \}$, and
	
\end{center}
%

\begin{center}
	$MTWL_{idf_{ij}}(a_k) = idf_{global}(a_k) \cdot idf_{local_{ij}}(a_k) \cdot f_{ij}(a_k)$
\end{center}

Next, \citeN{Muhr:2010} integrated sibling information as a weighting factor. The main idea is to select as labels those terms which occur often in one sibling cluster only. This scheme is called Inverse Cluster Weight Labeling (ICWL), and it is calculated by the inverse cluster frequency ($icf$). The $icf$ for the $k^{th}$ term in cluster $n_{ij}$ with parent cluster $n_i$ is defined as:
\begin{center}
	
	$icf_{ij}(a_k) = \exp \left( \frac{ \#(a_k , D_{n_{ij}}) }{ \left| D_{n_{ij}} \right| } \right) \log \left( \frac{ \#\left(n_{i}\right) }{ \#(a_k , n_{i}) } + 1 \right)$
	
\end{center}

\noindent where $\#(a_k , D_{n_{ij}})$ is the number of documents in the cluster $n_{ij}$ containing term $a_k$, $\left| D_{n_{ij}} \right|$ is the number of documents in the cluster $n_{ij}$, $\#\left(n_{i}\right)$ is the number of direct sub-clusters in the parent cluster $n_i$ of cluster $n_{ij}$ and $\#(a_k , n_{i})$ is the number of direct $n_i$ sub-clusters which contain the term $a_k$. The exponential component promotes terms occurring in a larger fraction of documents. The scheme is referred to as \textbf{ICWL$_{raw}$} when $icf$ weight is combined with MTWL$_{raw}$, and the labeling function is defined as:

\begin{center}
	$L(n_{ij}) = \{ best_p(A_{ij}), p=1,...,P \}$, where
\end{center}
%

\begin{center}
	$A_{ij} = \{ \forall a_k \in A \mid ICWL_{raw_{ij}}(a_k) > 0 , k=1,..,m \}$, and
	
\end{center}
%

\begin{center}
	$ICWL_{raw_{ij}}(a_k) = icf_{ij}(a_k) \cdot f_{ij}(a_k)$
\end{center}

The scheme is referred to as \textbf{ICWL$_{idf}$} in the case of $icf$ weight being combined with $MTWL_{idf}$. Then, the labeling function is defined as:

\begin{center}
	$L(n_{ij}) = \{ best_p(A_{ij}), p=1,...,P \}$, where
\end{center}
%

\begin{center}
	$A_{ij} = \{ \forall a_k \in A \mid ICWL_{idf_{ij}}(a_k) > 0 , k=1,..,m \}$, and
	
\end{center}
%

\begin{center}
	$ICWL_{idf_{ij}}(a_k) = idf_{global}(a_k) \cdot idf_{local_{ij}}(a_k) \cdot icf_{ij}(a_k) \cdot f_{ij}(a_k)$
\end{center}

Finally, \citeN{Muhr:2010} extended the labeling approaches by weighting the influence of a term into the path length of each descendant cluster of cluster $n_{ij}$. The main idea is to explore the ancestor-descendant relationship. According to the authors, incorporating sibling information may increase the discriminatory importance of parent labels that are equally distributed possibly over all siblings, and can decrease the importance of parent labels occurring often in few descendant clusters.

\begin{figure}
	\centering
	\includegraphics[width=0.6\textwidth]{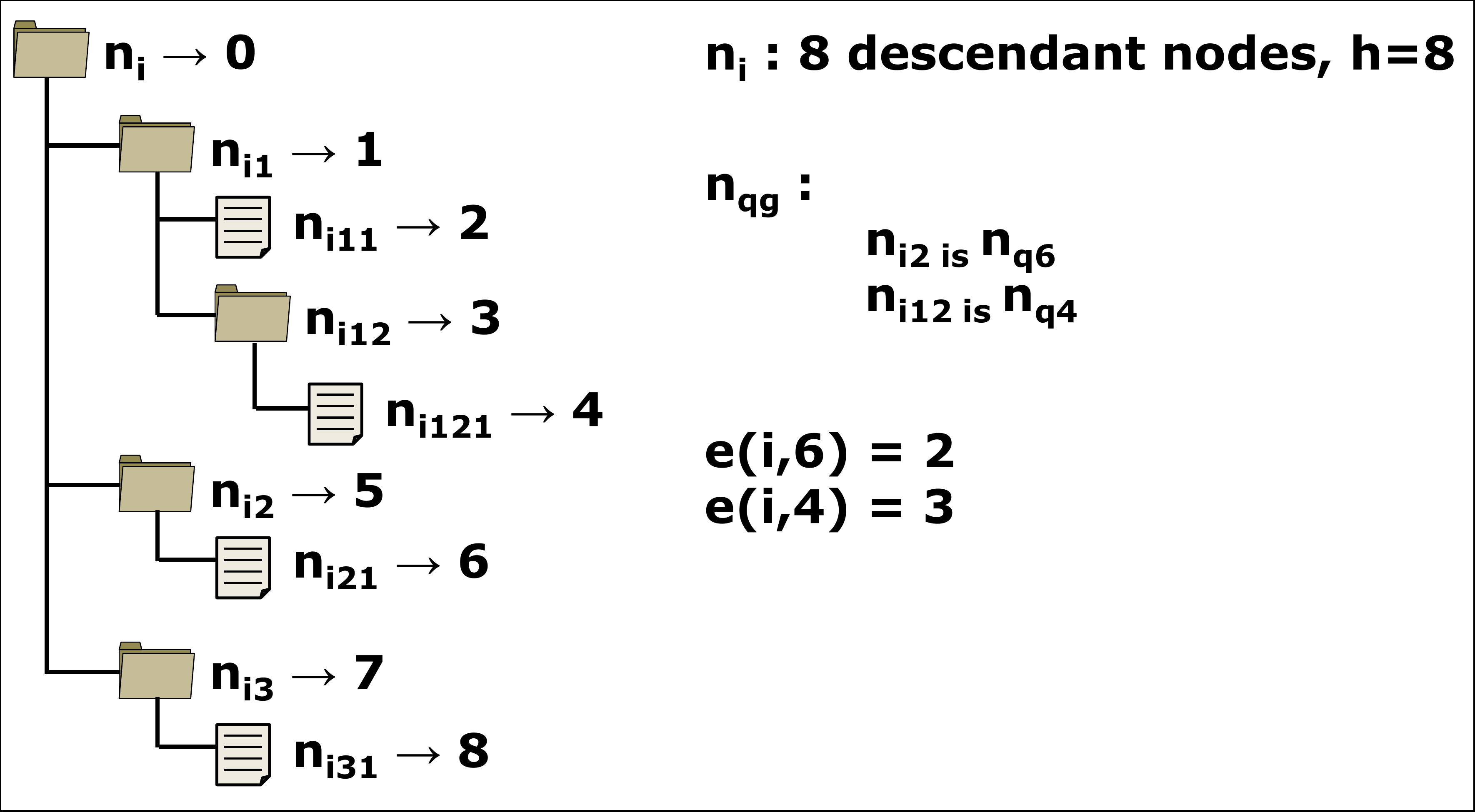}
	\caption{Descendant nodes example.}
	\label{fig:Descendant}
\end{figure}

Methods involving hierarchical weighting are prefixed here with ``Hier''. For each term $a_k$ in cluster $n_{i}$, a weight $w$ is associated. In order to define this weight, we first need to introduce some important definitions and an abstract example of them in Figure \ref{fig:Descendant}:

\begin{itemize}
	\item $n_{ig}$ ($g=1,...,h$) is a descendant node of $n_i$, that is, not only the direct children nodes.
	
	\item $n_{qg}$ is the direct parent of node $n_{ig}$.
	
	\item $e\left(i,g\right)$ is the path length between two nodes. The path length is the number of edges between n$_{i}$ and n$_{ig}$.
	
	\item $cf_{e\left(ig\right)}$ is the sibling base cluster frequency of a term, where:
	
	\begin{itemize}
		\item $cf_{e\left(ig\right)}(a_k) = \frac{ \#(a_k, n_{qg})}{ \#\left(n_{q}\right) }$, $\#(a_k, n_{qg})$ is the number of direct $n_{qg}$ sub-clusters of $n_q$ which contain $a_k$, $n_q$ is $n_{qg}$ father and $\#\left(n_{q}\right)$ is the $n_q$ number of children. 
	\end{itemize}
\end{itemize}

So, the weight $w$ for the $k^{th}$ term in cluster $n_{i}$ is defined as:

\begin{center}
	
	$w_{i}(a_k) = \displaystyle\sum\limits^{h}_{g=1} \frac{1}{e\left(i,g\right)} \cdot cf_{e\left(ig\right)}(a_k) \cdot v_{i,g}(a_k)$
	
\end{center}

\noindent where $v_{i,g}(a_k)$ is the weight of term $a_k$ in node $n_{ig}$ weighted in the local context of the node $n_i$.
The $v_{i,g}(a_k)$ is calculated according to the selected scheme, as presented bellow:

\begin{itemize}
	\item \textbf{HierMTWL$_{raw}$}:
	
	\subitem $w_{i}(a_k) = \displaystyle\sum\limits^{h}_{g=1} \frac{1}{e\left(i,g\right)} \cdot cf_{e\left(ig\right)}(a_k) \cdot f_{ig}(a_k)$
	
	\item \textbf{HierMTWL$_{idf}$}:
	
	\subitem $w_{i}(a_k) = \displaystyle\sum\limits^{h}_{g=1} \frac{1}{e\left(i,g\right)} \cdot cf_{e\left(ig\right)}(a_k) \cdot idf_{global}(a_k) \cdot idf_{local_{i}}(a_k) \cdot f_{ig}(a_k)$
	
	\item \textbf{HierICWL$_{raw}$}:
	
	\subitem $w_{i}(a_k) = \displaystyle\sum\limits^{h}_{g=1} \frac{1}{e\left(i,g\right)} \cdot cf_{e\left(ig\right)}(a_k) \cdot icf_{i}(a_k) \cdot f_{ig}(a_k)$
	
	\item \textbf{HierICWL$_{idf}$}:
	
	\subitem $w_{i}(a_k) = \displaystyle\sum\limits^{h}_{g=1} \frac{1}{e\left(i,g\right)} \cdot cf_{e\left(ig\right)}(a_k) \cdot idf_{global}(a_k) \cdot idf_{local_{i}}(a_k) \cdot icf_{i}(a_k) \cdot f_{ig}(a_k)$
	
\end{itemize}

\noindent and the labeling function for all cases is defined as:

\begin{center}
	$L(n_{i}) = \{ best_p(A_{i}), p=1,...,P \}$, where
\end{center}
%

\begin{center}
	
	$A_{i} = \{ \forall a_k \in A \mid w_{i}(a_k) > 0 , k=1,..,m \}$
\end{center}

For all cases, since the root node does not have a parent node, we considered the root node as a parent node of itself in this paper.

%
%
%
 
\subsection{Methods based on probabilistic or statistic models}
\label{sec:methods-probstat}

\citeN{Glover:2002} proposed a model exclusively based on the observed frequencies for each term in the group, as illustrated in Table \ref{ContTable} and Figure \ref{fig:ExemploTabela3}. In other words, the model is based on the maximum likelihood estimates of the conditional probabilities of the term ($a_k$ attribute) frequency in each \textit{$n_{ij}$} group, that is $f(a_k|n_{ij})$ and in the $a_k$ total frequency in the group, that is $f_i(a_k)$. Thus, the maximum likelihood estimators are $f_i(a_k)=\sum_j f_{ij}(a_k)$ for $a_k$ frequency in $n_i$ and $f(a_k|n_{ij})= f_{ij}(a_k)/\sum_j(\sum_{k}f_j(a_k))$ for the conditional probability of $a_k$ in $n_{ij}$. If $f(a_k|n_{ij})$ is very common and $f(a_k)$ is rare then the term is a good discriminator for the \textit{$n_{ij}$}, otherwise if $f(a_k|n_{ij})$ and $f(a_k)$ are common then the term discriminates the \textit{parent node of $n_{ij}$} ($n_i$) and, finally, if $f(a_k|n_{ij})$ is very common and $f(a_k)$ is relatively rare in the collection then the term is a better discriminator for the \textit{child node of $n_{ij}$}. The thresholds \textit{very common} and \textit{rare} must be ex\-pe\-ri\-men\-tal\-ly determined. 

The methods proposed by \citeN{PopeUngar:2000} and RLUM \cite{MouraRezende2010} are inspired in Glover's model,  based on a multinomial distribution of the attributes which is, in its turn, based on the obtained clusters. The label selection of a generic node is carried out by testing each attribute distribution dependence on the n$_{i}$ child nodes.

Consider Figure \ref{fig:ExemploTabela3} and the contingency Table \ref{ContTable}, and $a_k$ as a generic attribute with a non-zero frequency in $n_i$. If the attribute distribution is independent of the $n_i$ children, $n_{ij}$, $j=1,...,c$, then the attribute ($a_k$) does not discriminate the child nodes, so it is a good discriminator for the parent node $n_i$. Otherwise, the attribute is a good discriminator for some of the child nodes, $n_{ij}$. Also consider the follow definitions:

\begin{center}

	$s = \displaystyle\sum\limits^{c}_{j=1}\sum\limits^{m}_{t=1} f_{ij}(a_t)= \sum\limits^{m}_{t=1} f_{i}(a_t)$
\end{center}

\begin{center}	
	$TP = f_{ij}(a_k)$,  $FN = \sum\limits^{m}_{t=1} f_{ij}(a_t) - TP$
\end{center}

\begin{center}	
	$FP = f_{i}(a_k) - TP$,  $TN = s - (TP + FN +FP)$
\end{center}

Under the hypothesis of independence, that is, the $k^{th}$ term ($a_k$) does not depend on the $n_i$ children, each $f_{ij}(a_k)$ is supposed to depend exclusively on the marginal frequencies; i.e., the expected value for each $f_{ij}(a_k)$ must be:
\[
E(f_{ij}(a_k))=f_i(a_k)\times\sum^{m}_{t=1}f_{ij}(a_t)\times\frac{1}{\sum\limits^{m}_{t=1}f_{i}(a_t)}
\]
  
To test the independence hypothesis in the proposal of \citeN{PopeUngar:2000}, a chi-square estimate is calculated and the result is compared to a tabled value with $c-1$ degrees of freedom (the number of node children subtracted by one).The chi-square estimate is given by:

\begin{center}

$\chi^2_{i}(a_k) = \frac{ \left( TP \times TN - FN \times FP\right)^2 \times s }{ \left( TP+FN \right) \times \left( FP+TN \right) \times \left( TP+FP \right) \times \left( FN+TN \right) }$

\end{center}

In the method proposed by \citeN{PopeUngar:2000}, the chi-square estimate is tested to each $a_k$, $k=1,...,m$, all over the hierarchy starting at the root node; i.e., $\forall\ n_i, \ i=1,...,|N|$, assuming the tree has $|N|$ nodes. For each $a_k$, the method decides to use it in the $n_i$'s label or in the child node labels along the hierarchy. As the chi-square is not a robust estimate for low frequency values (see \cite{Bishop:1975}), this method assumes the most used rule of only applying this test when the attribute $a_k$ has $f_{ij}(a_k)\geq5$ in all $n_{ij}$ nodes. Consequently, this method does not make a decision for all $a_k$ attributes in all $n_i$ nodes, which sometimes causes term repetitions along the hierarchy. Besides, this method has a high computational complexity close to $O(m\times N^2)$, considering $m$ attributes and $|N|$ nodes.

Therefore, starting from the root node, the \textbf{Popescul\&Ungar} labeling function is defined as:

\begin{center}
   $L(n_{i}) = \{ best_p(A_{i}), p=1,...,P \}$, where
\end{center}
%

\begin{center}

  $A_{i} = \{ \forall a_k \in A \mid  \left( MTWL_{raw_{i}}(a_k \mid Popescul\&Ungar_{i}(a_k)\ is\ True)\ >0 \right), k=1,..,m \}$,
\end{center}

\noindent and
\[	Popescul\&Ungar_i(a_k) = \left(\ I\  is\  True\ \wedge \ II\  is\  True \wedge \ III\  is\  True \right) \]
\[  	I: a_k\ isn't\ selected\ as\ label\ of\ any\ ancestor\ of\ n_i  \] 
\[  	II:f_{ij}(a_k)\ \geq\ 5,\ \forall\ j=1,...,c  \] 
\[   	III: E(f_{ij}(a_k)) = f_i(a_k)\times\sum^{m}_{t=1}f_{ij}(a_t)\times\frac{1}{\sum^{m}_{t=1}f_{i}(a_t))} \]	
\[ \forall \ j=1,...,c \]

\noindent The condition \textit{I} ensures that each term is selected as label for only one node. However, this restriction does not prevent the same label from being used in nodes of a same level or in different branches of the hierarchy. This method does not make any decision for the leaves.

The Robust Labeling Up Method, RLUM \cite{MouraRezende2010}, test the expected frequency value of each $a_k$, $k=1,...,m$, all over the hierarchy, starting at the leaves. Besides the fact that it has a linear computational complexity, $O(m\times N)$, it deals with the test restrictions in a better way than Popescul\&Ungar and provides an embedded tree pruning.

To deal with the frequency estimate restrictions, firstly RLUM considers only the $a_k$ attributes which have $f_{ij}(a_k) \neq 0, \:j=1,...,c$ in each $n_{ij}$ child. This prevents the lack of decision when an attribute is completely associated to a specific $n_{ib}$ node, $b=1,...,c$; that is, $\forall j\neq b\ f_{ij}(a_k)=0 \ and \ f_{ib}(a_k)\neq0$. Furthermore, the independence test is carried out from a chi-square estimate based on a relaxed criterion (also in \cite{Bishop:1975}). This criterion is: $if\ f_{ij}(a_k)\geq1 \wedge \exists f_{ib}(a_k)\ggg 1 \ for\ some \ b\neq j \wedge \ b\in\{1,2,...,c\}\ \Rightarrow \chi^2 \ is \ a\ valid\ estimate$. As document clustering is a text mining problem, usually the frequency values for the attributes are very large, although some of these frequency values can be zero or very close to zero, because we are dealing with sparse matrix problems; consequently, this criterion can be applied. 

Thus, the \textbf{RLUM} labeling function is applied from the leaves to the root as:

\begin{center}
   $L(n_{i}) = \{ best_p(A_{i}), p=1,...,P \}$, where
\end{center}
%

\begin{center}

  $A_{i} = \{ \forall a_k \in A \mid  \left( MTWL_{raw_{i}}(a_k \mid RLUM_{i}(a_k)\ is\ True)\ >0 \right), k=1,..,m \}$, and
\end{center}

\[	RLUM_i(a_k) = \left(\ I\  is\  True\ \wedge \ II\  is\  True \right) \]
\[  	I:f_{ij}(a_k)\ > \ 0,\ \forall\ j=1,...,c  \] 
\[   	II: E(f_{ij}(a_k)) = f_i(a_k)\times\sum^{m}_{t=1}f_{ij}(a_t)\times\frac{1}{\sum^{m}_{t=1}f_{i}(a_t))} \]	

Additionally, as soon as the $L(n_i)$ is complete, all $L(n_{i,j})$ (descendant $n_i$ node labels) are updated: 

\[	L(n_{ij}) = L(n_{ij}) - [L(n_i)\bigcap L(n_{ij})] , \forall j=1,...,c \]

That is, if the term a$_k$ is selected as a label of n$_i$, then this term is removed from all n$_i$ children. 

Finally, the idea of tree pruning embedded in RLUM algorithms is based on the cutting of the nodes with empty label sets. If, after the tests and selections, a $L(n_{ij})$ results in an empty set, the $n_{i}$ node (parent node of $n_{ij}$) absorbs the $n_{ij}$ node, which is removed from the tree. In this case, the $n_i$ parent node inherits the child set nodes of $n_{ij}$.
 
The methods proposed by \citeN{PopeUngar:2000} as well as RLUM are based on the concept of term propagation in a topic taxonomy, as described in \citeN{Kashyap:2005}. In this concept, the generic labels of the node $n_i$ are formed by the union of its ancestor node labels and its own labels (specific labels). In other words, it is important to avoid unnecessary repetitions along the document hierarchy, also keeping the most generic terms in the highest levels and the most specific terms in the lower levels.

\citeN{Muhr:2010} also evaluated label selection methods based on statistical models. The authors called these approaches as Reference Collection based Labeling (RCL) because they are comparing the distribution of terms contained in clusters to terms contained in a reference collection of documents. They evaluated three comparative statistics, namely the Jensen-Shannon Divergence \cite{Carmel:2006}, Information Gain and $\chi^2$. The $P$ terms with the best test values are taken as labels. In our work, two statistics were selected for evaluations: Jensen-Shannon Divergence (JSD) and $\chi^2$. For each node n$_{ij}$, the reference collection is defined as all documents belonging to the cluster sub-tree of its direct parent n$_i$ excluding all documents contained in n$_{ij}$. Now, consider the follow definitions:
 
\begin{center}
	$s = \displaystyle\sum\limits^{m}_{t=1} f_{i}(a_t) - \displaystyle\sum\limits^{m}_{t=1} f_{ij}(a_t)$,   $TP = f_{ij}(a_k)$,   $FN = \sum\limits^{m}_{t=1} f_{ij}(a_t) - TP$
\end{center}

\begin{center}	
	$FP = (f_{i}(a_k) - f_{ij}(a_k)) - TP$,   $TN = s - (TP + FN + FP)$
\end{center} 

The scheme is named by \citeN{Muhr:2010} as RCL$\chi^2$ when $\chi^2$ is selected. Formally, the \textbf{RCL$\chi^2$} labeling function is defined as:

\begin{center}
   $L(n_{ij}) = \{best_p(A_{ij}), p=1,...,P\}$, where
\end{center}
%

\begin{center}

  $A_{ij} = \{ \forall a_k \in A \mid RCL\chi^2_{ij}(a_k) > 0 , k=1,..,m \}$, and
\end{center}
%

\begin{center}

	$RCL\chi^2_{ij}(a_k) = \frac{ \left( TP \times TN - FN \times FP\right)^2 \times s }{ \left( TP+FN \right) \times \left( FP+TN \right) \times \left( TP+FP \right) \times \left( FN+TN \right) }$

\end{center}

The scheme is named by \citeN{Muhr:2010} as RCLJSD when JSD is selected. Formally, the \textbf{RCLJSD} labeling function is defined as:

\begin{center}
   $L(n_{ij}) = \{best_p(A_{ij}), p=1,...,P\}$, where
\end{center}
%

\begin{center}

  $A_{ij} = \{ \forall a_k \in A \mid RCL\chi^2_{ij}(a_k) > 0 , k=1,..,m \}$, and
\end{center}
%

\begin{center}

	$RCLJSD_{ij}(a_k) = \left( \frac{TP}{TP+FN} \right) \times \log_2\left( \frac{TP}{TP+FN} \right) -
	\left( \frac{TP}{TP+FN} \right) \times \log_2\left( 0,5 \times \left( \frac{TP}{TP+FN} + \frac{TP + FP}{TP+FP+FN+TN} \right) \right) + 
	\frac{TP + FP}{TP+FP+FN+TN} \times \log_2\left( \frac{TP + FP}{TP+FP+FN+TN} \right) -
	\frac{TP + FP}{TP+FP+FN+TN} \times \log_2\left( 0,5 \times \left( \frac{TP}{TP+FN} + \frac{TP + FP}{TP+FP+FN+TN} \right) \right)
	$

\end{center}

\citeN{Muhr:2010} evaluated the RCL scheme combined with Hier weight as presented in the previous section. In this case, the weight $w$ for the $k^{th}$ term in cluster $n_{ij}$ was defined for the RCL$\chi^2$ scheme as \textbf{HierRCL$\chi^2$}, where:

\begin{itemize}
		\item $n_{ij(g)}$ ($g=1,...,h$) is a descendant node of $n_{ij}$, that is, not only the direct children nodes.
	
		\item $n_{qg}$ is the direct parent of node $n_{ij(g)}$.
		
		\item The weight $w_{ij}(a_k)$ is:
\begin{center}		
		\[	s = \displaystyle\sum\limits^{m}_{t=1} f_{i}(a_t), \ \ \ TP = f_{ij(g)}(a_k), \ \ \ FN = \displaystyle\sum\limits^{m}_{t=1} f_{ij(g)}(a_t) - TP \]
	
		\[	FP = (f_{qg}(a_k) - f_{ij(g)}(a_k)) - TP, \ \ \ TN = s - (TP + FN + FP) \]
	
		\[	w_{ij}(a_k) = \displaystyle\sum\limits^{h}_{g=1} \frac{1}{e\left(ij,g\right)} \cdot cf_{e\left(ij(g)\right)}(a_k) \cdot RCL\chi^2_{ij(g)}(ak) \]
\end{center}
\end{itemize}

The weight $w$ for the $k^{th}$ term in cluster $n_{ij}$ was defined for the RCLJSD scheme as \textbf{HierRCLJSD}, where:

\begin{itemize}
		\item $n_{ij(g)}$ ($g=1,...,h$) is a descendant node of $n_{ij}$, that is, not only the direct children nodes.
	
		\item $n_{qg}$ is the direct parent of node $n_{ij(g)}$.
		
		\item The weight $w_{ij}(a_k)$ is:
\begin{center}	
		\[ s = \displaystyle\sum\limits^{m}_{t=1} f_{i}(a_t), \ \ \ TP = f_{ij(g)}(a_k), \ \ \ FN = \displaystyle\sum\limits^{m}_{t=1} f_{ij(g)}(a_t) - TP \]
	
		\[ FP = (f_{qg}(a_k) - f_{ij(g)}(a_k)) - TP, \ \ \ TN = s - (TP + FN + FP) \]
	
		\[ w_{ij}(a_k) = \displaystyle\sum\limits^{h}_{g=1} \frac{1}{e\left(ij,g\right)} \cdot cf_{e\left(ij(g)\right)}(a_k) \cdot RCLJSD_{ij(g)}(ak) \] 
\end{center}
\end{itemize}

\noindent and the labeling function for HierRCL$\chi^2$ and HierRCLJSD is defined as:

\begin{center}
   $L(n_{ij}) = \{best_p(A_{ij}), p=1,...,P\}$, where
\end{center}
%

\begin{center}

  $A_{ij} = \{ \forall a_k \in A \mid w_{ij}(a_k) > 0 , k=1,..,m \}$
\end{center}

For all cases, since the root node does not have a parent node, we considered the root node as a parent node of itself in this paper.

In the approach proposed by \citeN{Lamirel:2008}, the traditional $recall$, $precision$ and \textit{F-measure} measures were redefined. For each term a$_k$ in the cluster n$_{ij}$, the \textit{Clustering Recall} (CRecall) represents the ratio between the frequency of a$_k$ in the cluster n$_{ij}$ and the frequency of a$_k$ in the whole collection:

\begin{center}

	\[ CRecall_{ij}(a_k) =  (f_{ij}(a_k))/(f(a_k)) \]

\end{center}

In the same way, the \textit{Clustering Precision} (CPrecision) represents the ratio between the frequency of a$_k$ in the cluster n$_{ij}$ and the sum of the frequencies of all terms in the cluster n$_i$:

\begin{center}

	\[ CPrecision_{ij}(a_k) = (f_{ij}(a_k))/(\displaystyle\sum\limits^{m}_{t=1} f_{ij}(a_t)) \]

\end{center}

With these measures, \citeN{Lamirel:2008} defined the Clustering F-Measure as:

\begin{center}

	\[ CFMeasure_{ij}(a_k) = \frac{ 2 \times CRecall_{ij}(a_k) \times CPrecision_{ij}(a_k) }{ CRecall_{ij}(a_k) + CPrecision_{ij}(a_k) } \]

\end{center}

\noindent CFMeasure is used as a basis for selecting the most suitable label sets from the clusters. \citeN{Lamirel:2008} proposed two strategies in order to select the labels: CFAverage and CFLeaveOneOut. In both strategies, the first step is to compute the CFMeasure of each term a$_k$ of each leaf node. Then, the $P$ top terms are selected as labels in each leaf node n$_{ij}$, that is:

\begin{center}
   $L(n_{ij}) = \{best_p(A_{ij}), p=1,...,P\}$, where
\end{center}
%

\begin{center}
  $A_{ij} = \{ \forall a_k \in A \mid CFMeasure_{ij}(a_k) > 0 , k=1,..,m \}$
\end{center}

\noindent After computing the CFMeasure in leaf nodes, the label selection method is applied from parent nodes to the root node.

In the \textbf{CFAverage} strategy, the CFMeasure of term a$_k$ in the cluster n$_i$ is estimated by the average CFMeasure value of this term in all direct $n_{ij}$ children nodes of n$_i$.

\begin{center}
   $L(n_{ij}) = \{best_p(A_{ij}), p=1,...,P\}$, where
\end{center}
%

\begin{center}
  $A_{ij} = \{ \forall a_k \in A \mid CFMeasure_{ij}(a_k) > 0 , k=1,..,m \}$, and
\end{center}
%

\begin{center}
	\[ CFMeasure_{i}(a_k) = (\displaystyle\sum\limits^{c}_{j=1} CFMeasure_{ij}(a_k))/(c)\]
\end{center}
\noindent where \textbf{\textit{c}} is the number of direct children of cluster n$_i$.

In the \textbf{CFLeaveOneOut} strategy, \citeN{Lamirel:2008} proposed a strategy analogous to the \textit{leave one out} solution used for learning method evaluations. The main idea of this strategy is to discriminate a cluster compared with other clusters in the same level. It is different from the other approaches that compare the node with the descendants and ancestors. To select the labels of $n_i$ node with direct child nodes $n_{ij}$ ($j= 1,...,c$), it creates a list with all clusters that are in the same level of the child nodes of $n_i$ including the node $n_i$. Next, all the $n_{ij}$ nodes are removed from this list. Suppose the level of $n_{ij}$ has Z nodes, the labeling function for the $k^{th}$ term in cluster $n_{i}$ is defined as:
\begin{center}
   $L(n_{ij}) = \{best_p(A_{ij}), p=1,...,P\}$, where
\end{center}
%

\begin{center}
  $A_{ij} = \{ \forall a_k \in A \mid CFMeasure_{ij}(a_k) > 0 , k=1,..,m \}$, and
\end{center}
%

\begin{center}

	\[ CRecall_{i}(a_k) = \frac{ f_i(a_k) }{ \displaystyle\sum\limits^{z}_{v=1} f_{iv}(a_k) - \displaystyle\sum\limits^{c}_{j=1} f_{ij}(a_k)}, \ \  CPrecision_{i}(a_k) = \frac{ f_{ij}(a_k) }{ \displaystyle\sum\limits^{m}_{t=1} f_{ij}(a_t) } \]
	\[ CFMeasure_{i}(a_k) = \frac{ 2 \times CRecall_{i}(a_k) \times CPrecision_{i}(a_k) }{ CRecall_{i}(a_k) + CPrecision_{i}(a_k) } \]

\end{center}

\section{Comparison Methodology}

As said before, a cluster labeling task independent of the applied clustering method can be viewed as an attribute selection problem. In this way, each cluster is considered as a "class". Following this idea,  the obtained document hierarchy (H) is assumed to reflect the classes of the documents through its nodes. Under this assumption and certain experimental conditions, we looked for measures which  could classify the obtained results for different labeling methods. The measures must give us some information about the labeling method habilities in capturing: cluster specific information; a better hierarchy comprehension at different levels of granularity; and, the most subjective coherent labels.

In this way, the comparison methodology encompasses five tasks as detailed in the next subsections: i) experimental data standardization and the hierarchical cluster (H) production, in order to apply different labeling methods to copies of H and consequently to make paired comparisons; ii) the construction of query expressions through different labels and hierarchical levels, in order to reflect the cluster label quality  at different levels of H granularity; iii) the selection of metrics to evaluate the quality of the retrieval process; iv) a reliable and paired statistical comparison among the metrics obtained after the retrieval process; and, v) considering each obtained label as a ``cluster topic'', the use of a metric which is supposed to simulate the human interpretation of the label coherence.

\subsection{Experimental data standardization}
\label{sec:Datastd}

The comparison methodology focuses on the evaluation of the labeling methods after a cluster hierarchy (H) is obtained. 
 Consequently, before we can apply this methodology, the following statements must be true or at least observed:
\begin{enumerate}
\item the text collection was preprocessed with a satisfactory method and was transformed in a attribute-value matrix. 
 The matrix can be indexed by term frequence ($tf$), $tf_idf$, or other measure required by the cluster algorithm;
\item a hierarchical document clustering algorithm was applied to the attribute-value matrix, the produced hierarchy (H) 
was evaluated and is ready to be used;
\item each labeling method will be applied to a H copy, in order to reach paired comparisons among node labels from different methods;
\item for all probabilistic labeling methods, it is necessary to set a common p-value. To reach a fair comparison among the different methods, we suggest a same conservative value to all p-values, for example $5\%$.
\item some labeling methods (for example, RLUM) could cause a transformation of the labeled hierarchy structure, due to its proper pruning process. For a fair comparison, we must avoid applying the pruning process of the labeling method. This statement guarantees the original hierarchy (H) is not changed and the applied label comparisons keep paired; and,
\item the obtained label sets from different labeling methods always have different cardinalities. Therefore, after the application of 
each label selection function, we select the best labels in all $n_i$ nodes to a maximum of ten ($P=10$) terms, 
that is $L(n_i) = \{best_p(A_i), p=1,...,10\}$.
\end{enumerate}

\subsection{Query expressions through cluster labels}
\label{sec:labelQuery}

\begin{figure}
\centering
\includegraphics[width=0.65\textwidth]{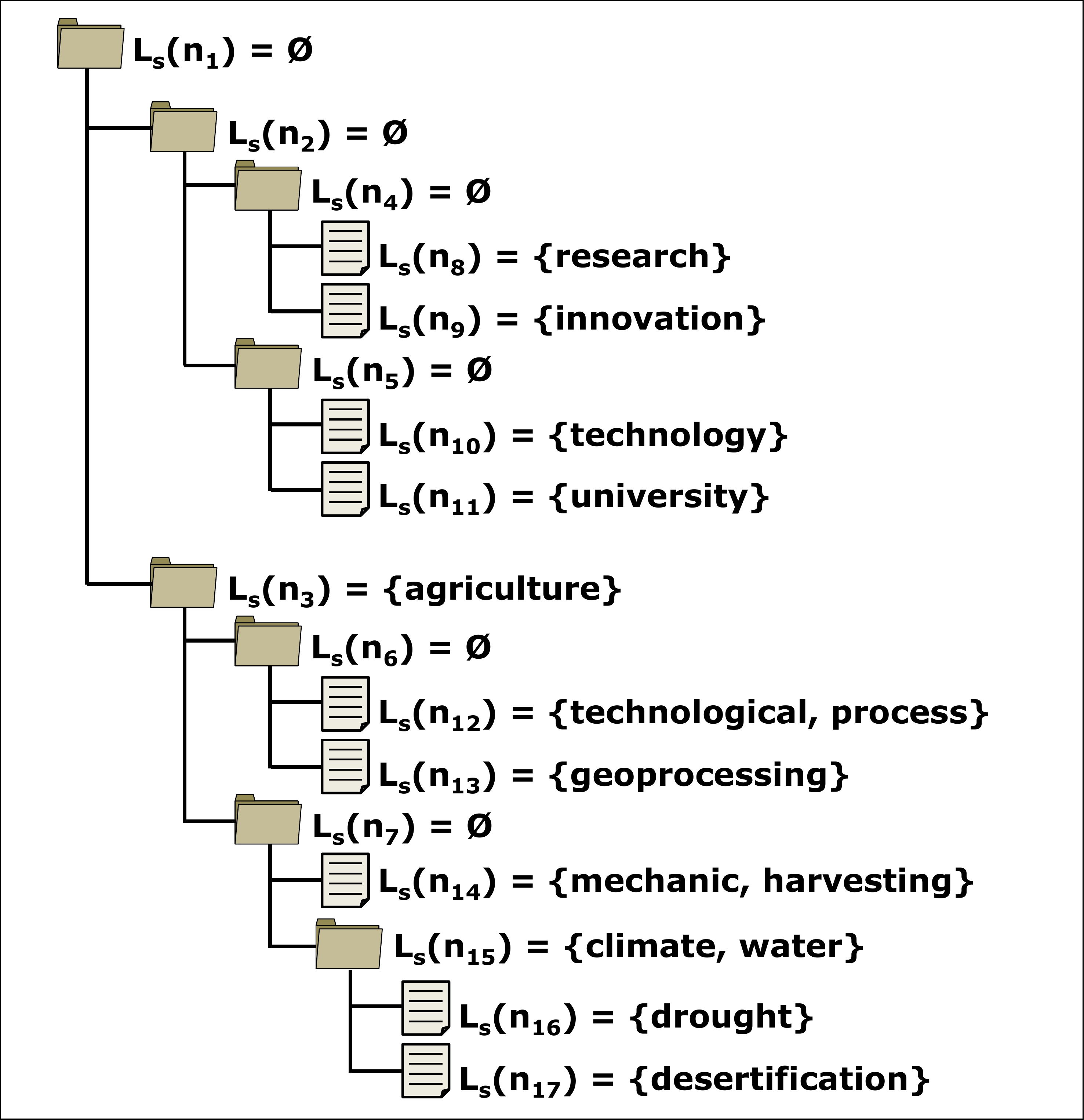}
\caption{A result example from a document cluster labeling.}
\label{fig:FIGX}
\end{figure}

A good hierarchical cluster labeling method has to be able to uniquely label each H node and to reflect the hierarchical relations among the nodes. Therefore, the most generic descriptors in a hierarchy are supposed to be in the highest nodes and the most specific in the leaves. For an example, in Figure \ref{fig:FIGX}, considering the groups (H nodes) as the actual classes of the documents in the groups and the obtained labels as the class descriptors, we can retrieve the documents under $n_3$ group using the query ``$agriculture$''. In a similar way, to retrieve the documents from $n_{12}$  we can use the query ``$technological\ OR\ process$'' or  ``($agriculture$)$\ AND \ ($technological\ OR\ process$)$''. In other words, as more specifically we construct the  query expression more restrictive are the obtained results. The construction of these query expressions allows to capture specific information in different levels of document hierarchy granularity.

To evaluate the results of query expressions, firstly we assume all $n_i$ labels ($L(n_i)$) as specific labels ($L_s(n_i)$). Thus, Figure \ref{fig:FIGX} presents an example of the specific labels ($L_s(n_i)$) of a hierarchical structure after applying a labeling method. Figure \ref{fig:FIGX} also presents some nodes which have empty specific labels. For example, RLUM can produce nodes with empty specific label sets when the label selection function is not true for all $a_k$. In these cases, RLUM performs a pruning process. As the pruning process is not allowed in this comparison methodology we must consider those empty label sets.

Also, we assume that the specific label set ($L_s(n_i)$) can be used to construct a query expression ($Q(L_s(n_i))$) where the elements in $L_s(n_i)$ are linked by an \textbf{OR} operator:

\begin{center}
$L_s(n_i)= \{a_{i1},a_{i_2},...a_{iP}\} \ \Rightarrow \ Q(L_s(n_i)) = [ a_{i1} \  OR \ a_{i_2} \  OR \ ... \  OR \ a_{iP} ]$
\end{center}

\begin{figure}
\centering
\includegraphics[width=0.85\textwidth]{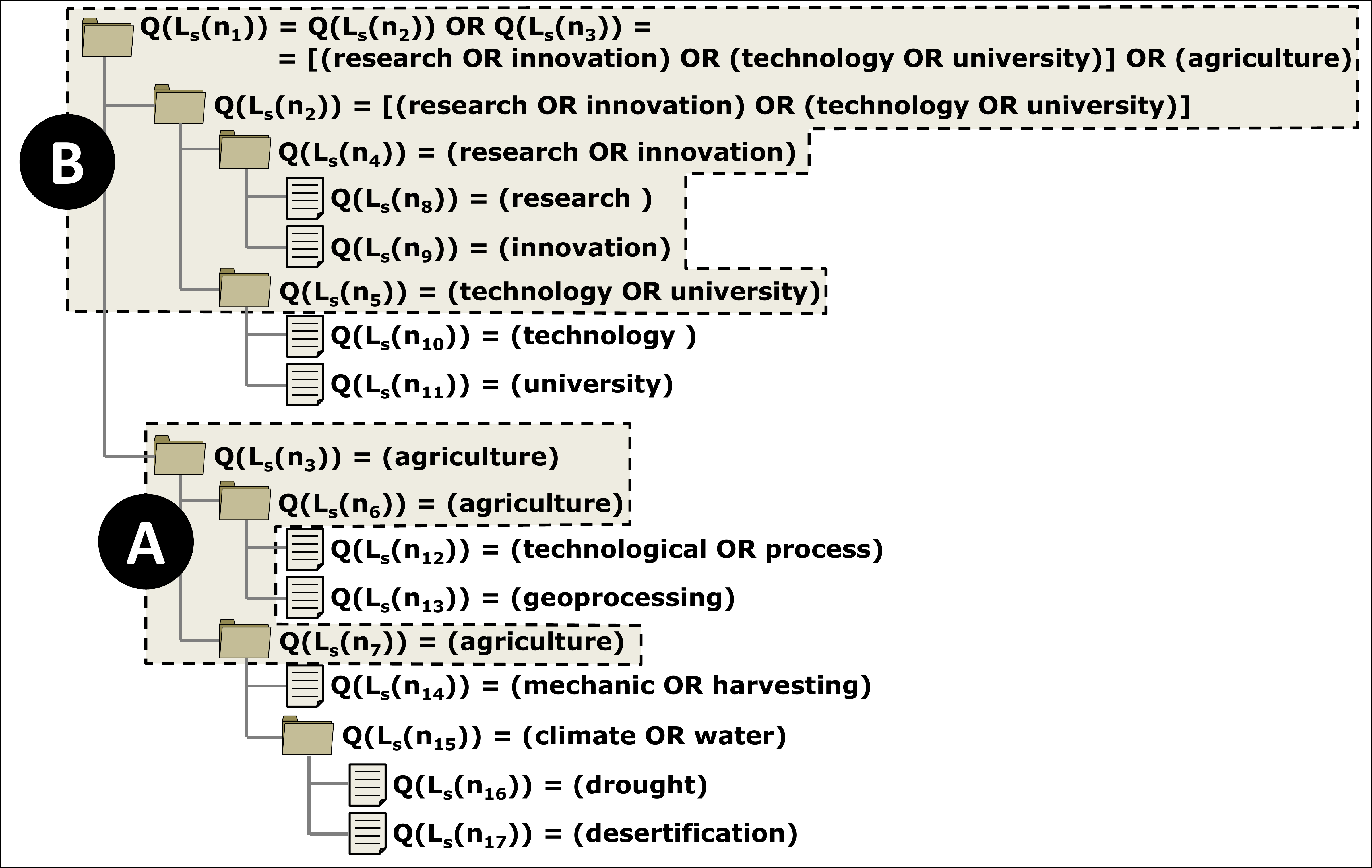}
\caption{The specific query expressions for the specific labels in  Figure \ref{fig:FIGX}.}
\label{fig:FIGW}
\end{figure}

The hierarchical structure can have nodes with an empty specific label which (i) has some ancestor with a non-empty label or (ii) has all ancestors with empty labels. In Figure \ref{fig:FIGX}, for example, there are $n_6$ and $n_7$ nodes with $n_3$ as a common ancestor. In order to have a correct interpretation of the document hierarchy, nodes in case (i) ($n_6$ and $n_7$) inherit the specific ancestor node label ($L_s(n_3)$). For example, in Figure~\ref{fig:FIGW}(A), $L_s(n_6)$ and $L_s(n_7)$ are interpreted as $L_s(n_3)$. 

Regarding (ii), those nodes have a lack of specific descriptors to compose their specific labels. Consequently, any of the descendent node labels could specify the node with the empty set. In other words, any of the children nodes are valid paths to retrieve the documents in the nodes with empty labels. So, the query expression for the empty sets can be the children query expressions linked by OR operators. For example, in Figure~\ref{fig:FIGW}(B) the query expressions for $n_5$, $n_4$, $n_2$ and $n_1$ are presented as:

\begin{center}

$Q(L_s(n_5))\ = \ Q(L_s(n_{10}))\ OR\ Q(L_s(n_{11}))\ =\ (technology)\ OR\ (university)\ =\ (technology\ OR\ university)$

$Q(L_s(n_4))\ = \ Q(L_s(n_8))\ OR\ Q(L_s(n_9))\ =\ (research)\ OR\ (innovation)\ =\ (research\ OR\ innovation)$

$Q(L_s(n_2))\ = \ Q(L_s(n_4))\ OR\ Q(L_s(n_5))\ =\ [(research\ OR\ innovation)\ OR\ (technology\ OR\ university)]$

$Q(L_s(n_1))\ = \ Q(L_s(n_2))\ OR\ Q(L_s(n_3))\ =\ [(research\ OR\ innovation)\ OR\ (technology\ OR\ university)]\ OR\ (agriculture)$

\end{center}

\begin{figure}
\centering
\includegraphics[width=0.85\textwidth]{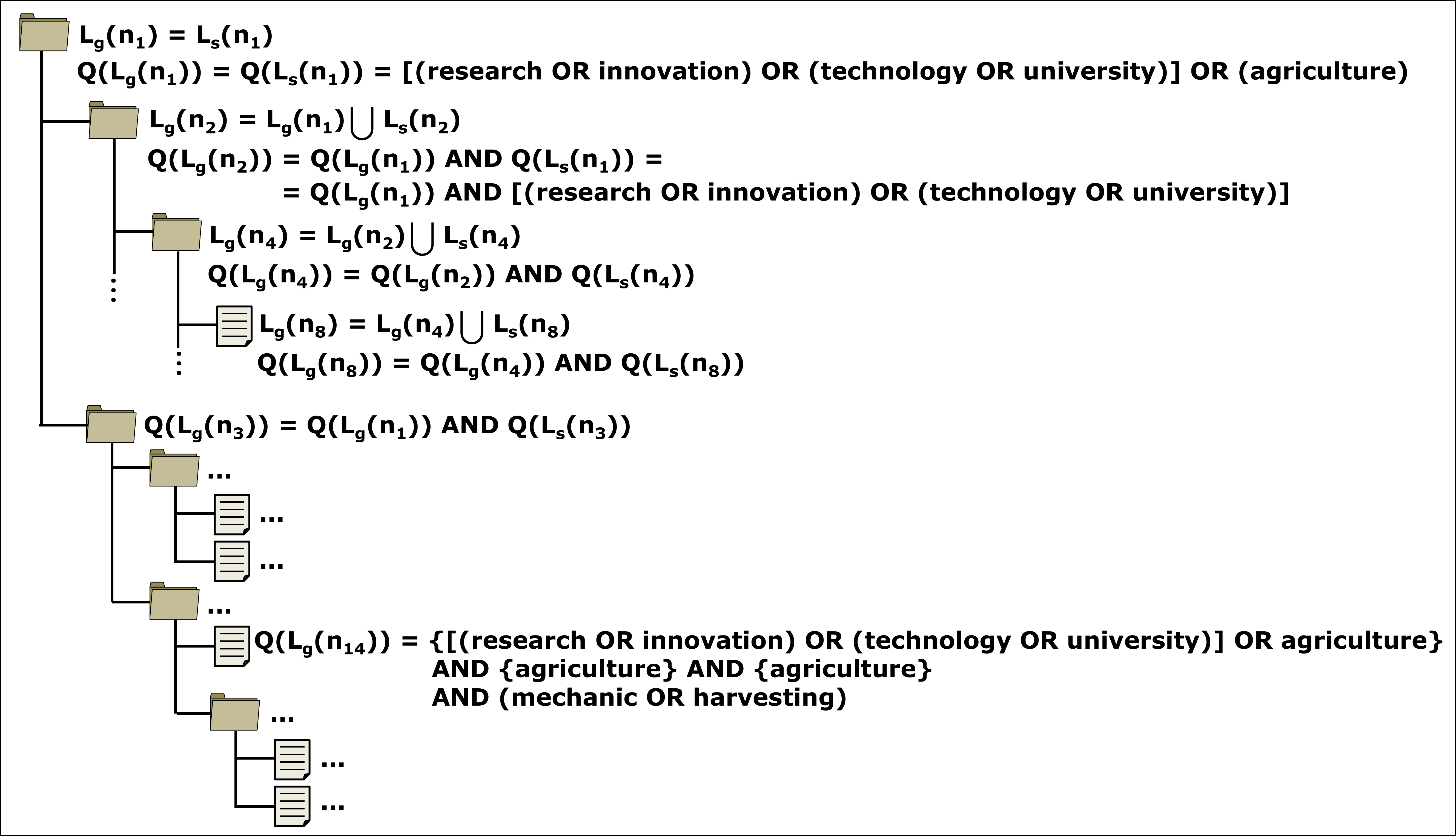}
\caption{The generic query expressions for the specific query expressions of  Figure \ref{fig:FIGW}.}
\label{fig:FIGZ}
\end{figure}

We must note that specific labels match specific nodes. That is, the specific labels ($L_s$) are supposed to describe the node as a flat (or isolated) cluster in the hierarchy. To reflect the hierarchical relations among the nodes, we need to specify generic labels ($L_g$), or hierarchical labels. In Figure \ref{fig:FIGZ}, the generic labels from Figure \ref{fig:FIGX} and their associated query expressions are illustrated.

A \textbf{generic label ($L_g$)} can be defined as a union of the specific label of a node ($n_i$) with its direct ancestor node labels. Therefore, the generic label for the hierarchy root node is the same as its specific label. The generic labels of nodes from the first level of the hierarchy are obtained by the union of their specific labels with the root node generic labels. Next, the generic labels of nodes from the second level are obtained by the union of their specific labels with the generic labels of their parent nodes. The same strategy is used in all nodes of the hierarchy. For the generic query expression ($Q(L_g(n_i))$), the operator link among the specific queries ($Q(L_s(n_i))$) of a node and the specific query of its ancestor node labels is \textbf{AND} in order to reflect the hierarchical dependence among them - as illustrated in Figure \ref{fig:FIGZ}.

\subsection{Metrics to evaluate the retrieval process}
\label{sec:RFP}  
 
In order to simulate the retrieval process, we: i) use the attribute-value matrix, which was generated to produce the H hierarchy, as the retrieval indexes; and, ii) build  query expressions as defined in the previous subsection. Thus, for a document to be retrieved, it must be indexed by the terms that satisfy the query expression. We are not considering the term frequency (or $tf_idf$)in the attribute-value matrix, but only whether this term belongs to the document or not. 
For example, in Figure \ref{fig:FIGX}, $L_s(n_{14})=\{mechanic,harvesting\}$ then the query $Q(L_s(n_{14}))= \ mechanic\ OR \ harvesting$ retrieves all documents which present the terms ``mechanic'' or ``harvesting''. 
In this way,  the retrieved documents which belong to the $n_{14}$ group are considered correctly retrieved; the incorrectly retrieved are the false positives, etc. Generalizing:

\begin{table}[htb] \centering
 \caption{Retrieval values - group (cluster).}
 \label{tab:measures}
 \small
 \begin{tabular}{c|c|c|c} 
               & $retrieved$ & $no\ retrieved$        &            \\ \hline
   $group$     & $t_p$       & $f_n$                 & $|g|$      \\ 
   $no\ group$  & $f_p$       & $t_n$                 &             \\  \hline
               & $|r|$       & $|\tilde{r}|$         &             \\                       
 \end{tabular}
\end{table} 

\begin{itemize}
\item $t_p$, or true positive: the number of documents that were retrieved and belong to the target group, or the correctly retrieved documents; 
\item $f_p$, or false positive: the number of documents that were retrieved and do not belong to the target group, or the incorrectly retrieved documents; 
\item $|r|$: the total number of retrieved documents;
\item $f_n$, or false negative: the number of documents in the target group which were not retrieved;
\item $t_n$, or true negative: the number of documents which were not retrieved and do not belong to the target group;
\item $|\tilde{r}|$ : the total number of documents that were not retrieved; and,
\item $|g|$ : the total number of documents in the target group.
 \end{itemize}
 
With the values in Table \ref{tab:measures}, we are able to obtain the following \textbf{measures} to evaluate the retrieval process:
 
 \begin{itemize}
\item \textbf{precision}: the proportion of correctly retrieved documents, $prec=\frac{t_p}{|r|}$;
\item \textbf{recall}, or retrieval coverage: the proportion of retrieved documents which were correctly retrieved and which were part of the target group, $rec=\frac{t_p}{|g|}$; and, 
\item \textbf{F$_{measure}$}: the harmonic mean between precision and recall, $F_{measure}=\frac{2*prec*rec}{prec+rec}$.
The ideal value for $F_{measure}$ is 1 because, for that to happen, the values of both precision and recall need to be 1.
 \end{itemize}
 
We must highlight the cases where the results produce recall or precision equal to zero. When this occurs, we are not calculating the \textbf{F$_{measure}$}; instead, we are assuming the harmonic mean is equal to the arithmetic mean, that is, zero - because zero documents were retrieved.
 
In order to experimentally evaluate the hierarchical clustering labeling methods, these measures are calculated for all labeling methods, specific query expressions ($Q(L_s(n_i))$) associated to each specific node label ($L_s(n_i)$) of the hierarchy and for each generic query expression ($Q(L_g(n_i))$) associated to each generic node label ($L_g(n_i)$).


\subsection{Pairwise comparisons of the obtained retrieval quality measures}
\label{sec:glm} 
The experimental results for each query expression quality measure ($m$, $m=$``precision'', ``recall'',``F'') are tabled for specific and generic query expressions, along with the number of the node ($n_i,\ i=1,...,o$), the cluster labeling method used to construct the node label ($l$, $l=$``MTWL$_{raw}$'', ``ICWL$_{raw}$'',..., see Table \ref{labelMethods}) and the hierarchical level number ($h,\ h=0,1,...$), in such a way that it is possible to generate its variance model.


We have sixteen different labeling methods, applied to hierarchical clusters. In this case, the variance analysis has to be able of evaluating the differences among the sixteen labeling methods, considering their effects  over different levels of the hierarchy. To achieve  a fair comparison, the total variance of a retrieval measure estimate has to be decomposed considering all those factors. This variance decomposition can be represented in this linear model (for details see \cite{Searle:1971}):

\begin{equation}
	\widehat{m(n_i)}\  = \ \hat{\mu} \  
	       	+\  \hat{h}\  +\   \hat{l}\   +\  \hat{\epsilon_{n_{i}}} 
	\label{glm:general}
\end{equation}

\begin{itemize}
 \item $\widehat{m(n_i)}$: estimated value for the retrieval measure in the $i_{th}$ node
  \item $\hat{\mu}$: estimated value for the general mean of the retrieval measure, considering the all other variance components were statistically decomposed 
 \item $\hat{h}$: estimated value for the hierarchical level component 
 \item $\hat{l}$: estimated value for the labeling method component 
 \item $\hat{\epsilon_{n_{i}}}$: estimated value of the model error component in the $i_{th}$  node, supposing it is random
\end{itemize}

This model adjustment allows the estimation of the deviations of the retrieval measure estimate ($\hat{m(n_i)}$) from its general mean ($\hat{\mu}$). It is important to note that each variance component has some different subcomponents. For example, the variance component from $\hat{l}$ is divided into sixteen subcomponents. The model adjustment can be obtained for a large number of statistical programs \footnote{For this paper we are using the general linear model from SOC, available on \url{http://repositorio.agrolivre.gov.br/}.}. After the adjustment,  we can analyze the importance of each subcomponent of the variance component through multiple mean comparisons. We are using one strong statistical multiple mean test, implemented in almost all the statistical software products - the SNK (Student-Newman-Keuls) test. 

In Table \ref{MMEANS}, we have to understand each labeling method effect ($\hat{l_1},\hat{l_2},\hat{l_3},\hat{l_4},...$) as the subcomponent of the variance component of the estimated labeling method $\hat{l}$. In other words, the labeling method causes a deviation in the estimated retrieval measure ($\widehat{m(n_i)}$). 
In Table \ref{MMEANS}, the variance deviation importances are grouped into $a,b,c,...$ groups. That is, the variance importance of the labeling methods $\hat{l_1}$ and $\hat{l_2}$ (in $a$ group) is statistically stronger than the  $\hat{l_3}$ (in $b$ group) and they are stronger than  $\hat{l_4}$ and $\hat{l_5}$ (in $c$ group), etc. In other words, the contributions of labeling methods $\hat{l_1}$ and $\hat{l_2}$ (in $a$ group) to the $\widehat{m(n_i)}$ are stronger than $\hat{l_3}$ contributions (in $b$ group), etc. Besides, the labeling methods in the same group have statistically the same deviation importance over the ($\widehat{m(n_i)}$) estimation. Furthermore, in Table \ref{MMEANS}, we have the measure estimate ($\widehat{m(n_i)}$) according to the adjusted model, that is, using the total error variance ($\widehat{V_r(E)}$)\footnote{The total error variance is the general mean error variance.}, its degrees of freedom ($df_r$) and a 5\% p-value ($\alpha$).   

\begin{table}[htb] \centering
 \caption{SNK - Multiple comparison of the mean values of the labeling method effects over $\widehat{m(n_i)}$.}
\label{MMEANS}
\small
  \begin{tabular}{c|c|c} 
   \multicolumn{3}{c}{$df_r=...$, $\widehat{V_r(E)}$=..., $\alpha$=0.05} \\
   $effect$  & $\widehat{m(l)}$   & $mean\ groups\ of\ the\ effects$  \\
   \hline
   $l_1$   & $\widehat{m(l_1)}$ & $a$ \\ 
   $l_2$   & $\widehat{m(l_2)}$ & $a$      \\
   $l_3$   & $\widehat{m(l_3)}$ & $\ \ b$      \\
   $l_4$   & $\widehat{m(l_4)}$ & $\ \ \ c$      \\
   $l_5$   & $\widehat{m(l_5)}$ & $\ \ \ c$      \\
    ...    & ...       & $...$     \\                             
 \end{tabular}
\end{table} 

In the same way, we can make multiple comparisons of means among the variance components through hierarchical levels ($\widehat{m(h)}$) over the measure estimate ($\widehat{m(n_i)}$), according to the adjusted model

Finally, assuming the level effects are in different \textit{mean groups of the effects}, it is interesting to evaluate the retrieval measures of each cluster labeling method among the hierarchical level effects. To achieve this comparison, we can use other general linear model, isolating the hierarchical effects. Thus, we can adjust the following general linear model:

\begin{equation}
	\widehat{m(h)}\  = \ \hat{\mu} \  +\  \hat{h}\   +\  \hat{\epsilon_{h}} 
	\label{glm:HieEffect}
\end{equation}

\begin{itemize}
 \item $\widehat{m(h)}$: estimated value for the retrieval measure in each $h$ level 
 \item $\hat{\mu}$: estimated value for the general mean of the retrieval measure in the $h$ level 
 \item $\hat{h}$: estimated value for the hierarchical level effect over the retrieval measure estimate
 \item $\hat{\epsilon_{h}}$: estimated value of the model error in the $h$ level 
\end{itemize}

With this adjusted model, for each hierarchical level effect, we can achieve the multiple mean comparisons among hierarchical levels. Although we have all the mean measures for each hierarchical level, the ranking of the level effects may not reflect the level sequence. In this case, to get a better idea of the effects over the levels, it is useful to plot the mean estimates in graphics $\widehat{m(h_k)}\times h_k,\forall k\ level$.

\subsection{Label interpretability evaluation}
\label{sec:topCoe}
In this evaluation, we are interested in the quality of the obtained labels through coherence or interpretability. Firstly it is assumed that the obtained labels reflect the topic of the group, that is the subject of the documents in the group.  \citeN{newmanTopics} propose a topic evaluation which they claim to agree with human interpretability. In the proposed evaluation, a set of terms selected as a topic is rated for a coherence measure using a reference collection. The coherence measure is estimated through the pointwise mutual information (PMI) between two different words. \citeN{lau2014machine} explored tasks of automatic evaluation for single topics and they showed that the observed coherence measure was able to emulate human performance. In addition to that, they improved the Newman formulation based on the normalized PMI (NPMI). The NPMI reduces the bias for PMI towards low frequency words and provides a standardized range ($\left[-1,1\right]$) for the calculated values. Besides, NPMI presented the best correlation values to human judgment in the experiments of \citeN{lau2014machine}. Therefore, in this paper we chose the coherence measure based on the pointwise mutual information between a pair $(a_{k_1},a_{k_2})$ in a $L(n_i)$, using the NPMI presented by \citeN{lau2014machine} 

\begin{center}
OC-NPMI$(L(n_i))= \displaystyle\sum\limits^{P}_{k_1=2}\sum\limits^{k_1-1}_{k_2=1}\frac{\log\left( 
\frac{P_{rob}(a_{k_1},a_{k_2})}{P_{rob}(a_{k_1}) \times P_{rob}(a_{k_2})}\right)}{-\log(P_{rob}(a_{k_1},a_{k_2}))} $
\end{center}

\noindent where, $P_{rob}$ is the probability function, and for each $L(n_i)$ all pairs of the $P$ top ranked attributes have their observed coherence (OC) calculated. That is, to calculate the observed co-occurrence as an estimate of the coherence through the NPMI, we use the observed frequencies of the attributes in the reference collection. The terms are taken in pairs and then their co-occurrences in the collection are calculated, always considering a sliding window of a predefined number of words. In this paper, the reference collections were the Wikipedia in English (Version date: $Aug\ 5th,\ 2013$) and in Portuguese (Version date: $Sep\ 2nd,\ 2013$). The coherence estimates were calculated using the topic-interpretability toolkit\footnote{Available on https://github.com/jhlau/topic\_interpretability/.}. Moreover, we considered  the sliding window size as each single text, because the number of single texts in both reference collections are statiscally large. Thus, the results of each cluster labeling methods were submitted to the observed coherence calculation.   

\begin{figure}[!h]
\centering
\includegraphics[width=1\textwidth]{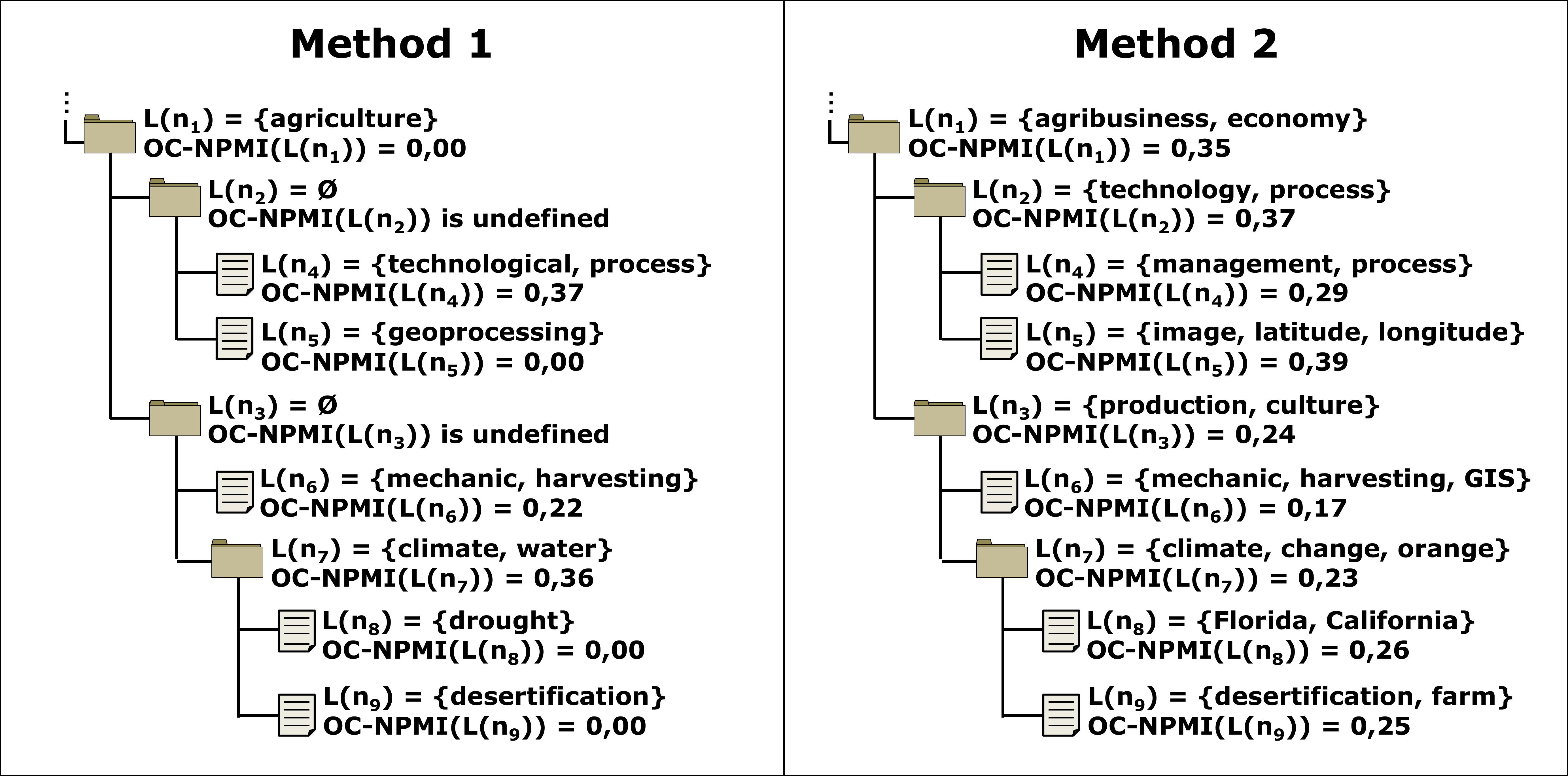}
\caption{An example of two different labeling method results and the observed coherence of their labels}
\label{fig:ArvoreExemploCoerenciaObservada}
\end{figure}

Consider the example in Figure \ref{fig:ArvoreExemploCoerenciaObservada} in which $Method\ 1$ and $Method\ 2$ are different cluster labeling methods applied to copies of a H. Performing a purely subjective analysis of the obtained labels, on average the $Method\ 2$ seems to select more consistent labels. Although, the $L(n_6)$ and $L(n_7)$ of $Method\ 2$ has specific terms (GIS and orange), which has undermined their interpretation. It should be noted that the $OC-NPMI$ calculated for each label reflects all these subjective observations - see Figure \ref{fig:ArvoreExemploCoerenciaObservada}. Additionally, in this work we considered all labels with a unique term have the average measure of $OC-NPMI$ - zero value.  

As the number of labels is usually huge, we are assuming that the labels  in the upper quartile of the coherence values are the most interesting. Also, we can use the observed coherence value which selects the best 25\% of nodes to compare label coherence among the cluster labeling methods.

Additionally, it is a \textit{condicio sine qua non} to apply the same information extraction process used in the preprocessing of the text collections to the reference collection, before the calculation of the coherence measures to each word pair. In this way, each attribute  matches a word in the reference collection.  

\section{Experimental set up}

We selected five text collections on which to apply the comparison methodology. All the text collections were submitted to the same data standardization process and then to the different labeling methods. Then, the proposed comparison methodology was applied to compare the topics produced by each method.

We assumed the preprocessing results summarized in Table \ref{TextCollections}. The first, second and third text collections cor\-respond to complete articles, in English, about chemistry, computer science and physics. The fourth text collection corresponds to articles in Portuguese from the Second IFM Assembly (IFM)\footnote{Instituto Fábrica do Milênio, IFM, is a Brazilian organization whose actions focus on solutions for the manufacturing industry.}. The second column in Table \ref{TextCollections} corresponds to the number of documents in each text collection\footnote{The first four text collections can be accessed from the author's project website: \url{http://sites.labic.icmc.usp.br/toptax/}.}. Additionally, to provide a blueprint of robustness of our proposed methodology in a big data scenario, we carried out an experiment with the Pubmed-Cancer text collection \cite{RossiRT:2013}. It is composed by abstracts of scientific articles about 12 types of cancer extracted from Pubmed\footnote{Available on \url{http://www.ncbi.nlm.nih.gov/pubmed}.}.  In the third column of Table \ref{TextCollections} there is the number of 1-gram attributes obtained from each text collection, for which the language stopwords were dropped (such as articles, interjections, numbers, etc.) and the Porter stemming process was applied \cite{porter:1997}\footnote{Available for Pretext2 tool, on \url{http://www.labic.icmc.usp.br/pretext2/}.}. In the fourth column there is Salton's $df$ filter; which considers $df$ as the document frequency of each attribute. Salton demonstrated that it is enough to use only the attributes which are present in at least $1\%$ and $10\%$ of the documents \cite{salton:1975}. Finally, in the last column, there is the cardinality of the attribute set A, which is how many attributes were used for each text collection.
 
\begin{table}[!ht]  \centering
 \caption{Number of documents and attributes in the text collections.}
 \label{TextCollections}
\small
  \begin{tabular}{|c|c|c|c|c|}
      \hline
      text collection        & \#docs   & \#1-grams & Salton's DF filter             & $card(A)$ \\ \hline
      Chemistry              &  328       & 49090        & $3 \leq df \leq 33$        &  9864        \\ \hline
      Computer Science  &  335       & 35855        & $3 \leq df \leq 34$        &  5955        \\ \hline
      Physics                  &  321       & 35447        & $3 \leq df \leq 32$         &  6425        \\ \hline
      IFM                        &  385       & 33058        & $4 \leq df \leq 39$         &  8509        \\ \hline
      PubMed-Cancer     &  65991   & 53443        & $660 \leq df \leq 6599$ &  1168        \\ \hline      
  \end{tabular}
\end{table}

After the preprocessing task, the text collections were represented in an attribute-value matrix, with each document in a line and the $A$ set in the columns. To index the attributes into the documents, we used the term frequency measure ($tf$). The first four  attribute-value matrices was submitted to the same clustering algorithm. We chose the average linkage algorithm because it is agglomerative, which is recommended for small text collections, and because it generally has the best performance among the agglomerative algorithms \cite{zhao2005hierarchical} with the cosine similarity measure; also, the R\footnote{The R project for statistical community, on http://www.r-project.org/.} implementation was used. On the other hand, the attribute-value matrix from PubMed-Cancer collection was  submitted to the bisecting K-means cluster method\footnote{Available on \url{http://sites.labic.icmc.usp.br/torch/msd2011/tophclust/}.} to extract its hierarchical structure. In this case, the bisecting k-means was more appropriated than the average, because of the document text dimension \cite{steinbach2000comparison}.

\begin{table}[!ht]  \centering
 \caption{Balance of resulting dendrogram according to the text collection.}
\label{tab:HierBalancing}
\small
       \begin{tabular}{|c|c|c|} \hline
                               &  \textbf{balanced}  & \textbf{unbalanced}  \\ \hline
      			 \textbf{text collection}   &  Physics     & Chemistry    \\ 
                               &  Computer & IFM               \\ 
                               &  PubMed-Cancer &                \\ \hline
      \end{tabular}
\end{table}

We did not apply any pruning or smoothing process on the obtained hierarchies. Consequently, there were balanced and unbalanced hierarchies, as summarized in Table \ref{tab:HierBalancing}. These effects were useful to the evaluation of results, as those hierarchies were submitted to the sixteen different cluster labeling methods and the proposed statistical evaluation.

\section{Results and Discussion}

%
In the Table \ref{labelMethods}, the cluster labeling methods used in this experiment are divided into two classes: cluster reference collection, and global reference collection. This classification's goal is to identify the methods which use a local cluster reference collection or the complete reference collection to aid in the labeling ranking. In the cluster reference collection methods, the ranking of labels in each cluster is defined based on the weight of the label in the cluster or in the direct parent cluster. The global reference collection methods include the weight of labels in the ancestors, descendants or in the text collection in some way to affect the labeling ranking.

\begin{table}[!ht]  \centering
 \caption{Label selection methods used in the experiment.} 
\label{labelMethods}
\resizebox{0.8\textwidth}{!} { 
\begin{tabular}{cc}

  \begin{tabular}{|c|c|}
      \hline
      \multicolumn{2}{|c|}{\textbf{Cluster reference collection}} \\ \hline
      \textbf{name}          &    \textbf{category}  \\ \hline
MTWL$_{raw}$        &    frequency based \\
\hline
ICWL$_{raw}$        &    frequency based \\
\hline
RCL$\chi^{2}$       &    probabilistic   \\
\hline
RCLJSD              &    probabilistic   \\
\hline
CFLeaveOneOut       &    probabilistic   \\
\hline
\end{tabular}
&
\begin{tabular}{|c|c|}
\hline
\multicolumn{2}{|c|}{\textbf{Global reference collection}} \\ \hline
      \textbf{name}          &    \textbf{category} \\ \hline
HierMTWL$_{raw}$    &    frequency based \\ 
\hline
MTWL$_{idf}$        &    frequency based \\
\hline
HierMTWL$_{idf}$    &    frequency based \\
\hline
ICWL$_{idf}$        &    frequency based \\
\hline
HierICWL$_{raw}$    &    frequency based \\ 
\hline
HierICWL$_{idf}$    &    frequency based \\
\hline
Popescul\&Ungar     &    probabilistic \\
\hline
RLUM                &    probabilistic \\
\hline
HierRCL$\chi^{2}$   &    probabilistic \\ 
\hline
HierRCLJSD          &    probabilistic \\ 
\hline
CFAverage           &    probabilistic \\
\hline
  \end{tabular}
  
  \end{tabular}}
\end{table}


Firstly, in this section, we present the multiple mean comparisons among the methods and hierarchical levels, using the specific and the generic search queries. Then, we make a careful analysis of the retrieval results among the hierarchical levels for each text collection. Additionally, the method's analysis is complemented by the observed coherence values of the obtained topics. Finally, there is a discussion on the data standardization and some methodology future work.
\vspace{-5mm}

\subsection{Retrieval measure variance results between methods}
\label{sec:Resglm}
%

The results in this section are from the use of the topics (label sets) as query expressions. We simulate the retrieval process in two steps, using the $Q(L_s)$ and $Q(L_g)$ for the resulting topics of each labeling method, and calculate the \textit{F measure} for the retrieval results (as proposed in section \ref{sec:RFP}). We are assuming that the \textit{F measure} is enough to observe the general results, because it is the harmonic mean of precision and recall. Then we can compare the \textit{F measure} means for each method through multiple mean comparisons, as explained in section \ref{sec:glm}.

\begin{table}[!ht] \centering
 \caption{Multiple mean comparisons for specific label queries ($Q(L_s)$) - by methods.}
\label{resultadoMetodos:Especificos}
\resizebox{0.8\textwidth}{!} {
 \begin{tabular}{|c|} \hline
   F measure\\ \hline
   \begin{tabular}{c||c}
          Chemistry  & Computer  \\
          (df=10412, $\widehat{V(E)}$=0.0089, $\alpha$=0.05)  & (df=10667, $\widehat{V(E)}$=0.0046, $\alpha$=0.05)     \\ \hline
          \begin{tabular}{c|c|l}
           method                                       & est.mean      & group \\ \hline
           ICWL$_{idf}$					& 0.086439	& $a$ \\
           MTWL$_{idf}$					& 0.084654	& $a$ \\
           Popescul\&Ungar			        & 0.083800	& $a\ b$ \\
           CFLeaveOneOut				& 0.078840	& $a\ b\ c$ \\
           RCL$\chi^{2}$				& 0.078010	& $a\ b\ c$ \\
           RCLJSD					& 0.077866	& $a\ b\ c$ \\
           ICWL$_{raw}$					& 0.076486	& $a\ b\ c$ \\
           MTWL$_{raw}$					& 0.075543	& $a\ b\ c$ \\
           HierRCL$\chi^{2}$		                & 0.073744	& $a\ b\ c$ \\
           HierMTWL$_{idf}$			        & 0.072061      & $a\ b\ c$ \\
           HierICWL$_{idf}$			        & 0.072061      & $a\ b\ c$ \\
           HierRCLJSD					& 0.071603      & $a\ b\ c$ \\
           HierICWL$_{raw}$			        & 0.070280      & $a\ b\ c$ \\
           HierMTWL$_{raw}$			        & 0.069155      & $a\ b\ c$ \\
           CFAverage				        & 0.066830      & $\ \ b\ c$\\
           RLUM						& 0.065548      & $\ \ b\ c$\\
          \end{tabular} 
         &
          \begin{tabular}{c|c|l}
           method & est.mean & group \\ \hline
	ICWL$_{idf}$                                    & 0.081279      &    $a$  \\
	MTWL$_{idf}$                                    & 0.080996      &    $a$  \\
	Popescul\&Ungar                                 & 0.077207      &    $a\ b$  \\
	CFLeaveOneOut                                   & 0.075540      &    $a\ b$  \\
	MTWL$_{raw}$                                    & 0.073650      &    $a\ b$  \\
	ICWL$_{raw}$                                    & 0.073599      &    $a\ b$  \\
	RCL$\chi^{2}$                                   & 0.073307      &    $a\ b$  \\
	RCLJSD                                          & 0.073307      &    $a\ b$  \\
	HierRCL$\chi^{2}$                               & 0.071953      &    $a\ b\ c$  \\
	HierMTWL$_{idf}$                                & 0.071151      &    $a\ b\ c$  \\
	HierICWL$_{idf}$                                & 0.071151      &    $a\ b\ c$  \\
	HierICWL$_{raw}$                                & 0.069411      &    $a\ b\ c$  \\
	HierMTWL$_{raw}$                                & 0.068416      &    $\ \ b\ c$  \\
	HierRCLJSD                                      & 0.066918      &    $\ \ b\ c$  \\
	CFAverage                                       & 0.065866      &    $\ \ b\ c$  \\
	RLUM                                            & 0.061223      &    $\ \ \ \ c$  \\   
          \end{tabular}  \\ \hline \hline
         IFM  & Physics  \\
          (df=11980, $\widehat{V(E)}$=0.0089, $\alpha$=0.05)  & (df=10217, $\widehat{V(E)}$=0.0075, $\alpha$=0.05)     \\ \hline
          \begin{tabular}{c|c|l}
           method & est.mean & group \\ \hline
	ICWL$_{idf}$ & 0.088404      &    $a$  \\
	Popescul\&Ungar & 0.086099   &    $a$  \\
	MTWL$_{idf}$ & 0.085376      &    $a$  \\
	RCLJSD & 0.080887            &    $a$  \\
	RCL$\chi^{2}$ & 0.080877     &    $a$  \\
	ICWL$_{raw}$ & 0.080267      &    $a$  \\
	MTWL$_{raw}$ & 0.079389      &    $a$  \\
	HierICWL$_{idf}$ & 0.078862  &    $a$  \\
	HierMTWL$_{idf}$ & 0.078862  &    $a$  \\
	RLUM & 0.076586              &    $a$  \\
	HierICWL$_{raw}$ & 0.076530  &    $a$  \\
	HierRCL$\chi^{2}$ & 0.075677 &    $a$ \\
	CFAverage & 0.075535         &    $a$  \\
	HierMTWL$_{raw}$ & 0.075505  &    $a$  \\
	CFLeaveOneOut & 0.074748     &    $a$  \\
	HierRCLJSD & 0.072860        &    $a$ \\
	  \end{tabular}
	  &
	  \begin{tabular}{c|c|l}
           method & est.mean & group \\ \hline
	ICWL$_{idf}$         & 0.090611 &    $a\ \ \ \ $  \\
	MTWL$_{idf}$         & 0.089944 &    $a\ \ \ \ $  \\
	Popescul\&Ungar      & 0.083274 &    $a\ b\ \ $  \\
	RCL$\chi^{2}$        & 0.080830 &    $a\ b\ c$  \\
	RCLJSD               & 0.080720 &    $a\ b\ c$  \\
	HierICWL$_{idf}$     & 0.079910 &    $a\ b\ c$  \\
	HierMTWL$_{idf}$     & 0.079910 &    $a\ b\ c$  \\
	ICWL$_{raw}$         & 0.079553 &    $a\ b\ c$  \\
	MTWL$_{raw}$         & 0.079213 &    $a\ b\ c$  \\
	HierICWL$_{raw}$     & 0.077328 &    $a\ b\ c$  \\
	HierRCL$\chi^{2}$    & 0.077200 &    $a\ b\ c$  \\
	CFLeaveOneOut        & 0.076857 &    $a\ b\ c$  \\
	HierMTWL$_{raw}$     & 0.076373 &    $a\ b\ c$  \\
	CFAverage            & 0.073346 &    $\ \ b\ c$  \\
	HierRCLJSD           & 0.073105 &    $\ \ b\ c$  \\
	RLUM                 & 0.066891 &    $\ \ \ \ c$  \\
	  \end{tabular}
      \end{tabular} \\ \hline
 \end{tabular}}
\end{table}

The \textit{F measure} adjusted means presented in Table \ref{resultadoMetodos:Especificos} were calculated from the retrieval results, using the specific query expressions ($Q(L_s)$) as specified in Section \ref{sec:glm}. We can observe from one to three $mean\ groups$ of \textit{F measure} means. However, the means are very close, because the deviations are very small ($\widehat{V(E)}$). Besides, a more detailed analysis of the results revealed that the \textit{F measure} means have very small values because of big recall values (close to one) and very small precision values (close to zero).

Using specific query expressions ($Q(L_s)$), as illustrated in Table \ref{resultadoMetodos:Especificos}, the methods $ICWL_{idf}$ and $MTWL_{idf}$ are always in the first mean group and they use a weight which combines a local and a global ``idf''. They are followed by the methods which use the cluster reference collection to aid label ranking and by the method proposed by Popescul\&Ungar. On the other hand, other methods present in the $a$ group use global reference collection to aid in label selection. We must note that the IFM collection, which has an unbalanced hierarchy, has only one $mean\ group$, that is, its results are not conclusive. For the Physics and Computer collections, methods such as RLUM, CFAverage and HierRCLJSD are more distant from the others. Even so, we can observe that there are no significant differences among the evaluated cluster labeling methods using the specific labels as query expressions. Statistically, they obtain similar results. We must note that a similar behavior is presented from the biggest text collection using specific label queries ($Q(L_s)$), in Table  \ref{F:Limitations}.  We can observe that  $MTWL_{idf}$ and $ICWL_{idf}$ are also among the first mean group. Although $Popescul\&Ungar$ had a little F value improvement.

Ainda, após a análise apresentada dos resultados da specific query expressions ($Q(L_s)$) evaluation, pode-se observar que os métodos que priorizam a seleção de termos específicos para cada grupo, visando diferenciar o grupo pai do grupo filho por esses termos, acabam se destacando. Assim, a avaliação de specific query expressions ($Q(L_s)$) permite a avaliação e escolha de métodos de seleção de descritores quando as características desejadas forem a de seleção de termos específicos por grupo considerando a hierarquia ``localmente'' (local = olha apenas a relação pai-filho).

\begin{table} \centering
 \caption{Multiple mean comparisons for generic label queries ($Q(L_g)$) - by methods.}
\label{resultadoMetodos:Genericos}
\resizebox{0.8\textwidth}{!} {
 \begin{tabular}{|c|} \hline
   F measure\\ \hline
   \begin{tabular}{c||c}
          Chemistry  & Computer  \\
          (df=10412, $\widehat{V(E)}$=0.0248, $\alpha$=0.05)  & (df=10667, $\widehat{V(E)}$=0.0280, $\alpha$=0.05)     \\ \hline
          \begin{tabular}{c|c|l}
           method & est.mean & group \\ \hline
						RCL$\chi^{2}$                & 0.138999 &    $a$  \\
						RCLJSD                       & 0.138786 &    $a$  \\
						RLUM                         & 0.117675 &    $\ \ b$  \\
						HierRCL$\chi^{2}$            & 0.108708 &    $\ \ b\ c$   \\
						MTWL$_{raw}$                 & 0.098216 &    $\ \ b\ c\ d$  \\
						MTWL$_{idf}$                 & 0.089902 &    $\ \ \ \ c\ d\ e$  \\
						HierICWL$_{raw}$             & 0.087184 &    $\ \ \ \ c\ d\ e$   \\
						HierMTWL$_{raw}$             & 0.077192 &    $\ \ \ \ \ \ d\ e\ f$ \\
						ICWL$_{raw}$                 & 0.071710 &    $\ \ \ \ \ \ \ \ e\ f\ g$ \\
						ICWL$_{idf}$                 & 0.064186 &    $\ \ \ \ \ \ \ \ \ \ f\ g$ \\
						Popescul\&Ungar              & 0.062783 &    $\ \ \ \ \ \ \ \ \ \ f\ g$  \\
						HierICWL$_{idf}$             & 0.052160 &    $\ \ \ \ \ \ \ \ \ \ \ \ \ g\ h$  \\
						HierMTWL$_{idf}$             & 0.052160 &    $\ \ \ \ \ \ \ \ \ \ \ \ \ g\ h$  \\
						HierRCLJSD                   & 0.038240 &    $\ \ \ \ \ \ \ \ \ \ \ \ \ \ \ h$  \\
						CFLeaveOneOut                & 0.034680 &    $\ \ \ \ \ \ \ \ \ \ \ \ \ \ \ h$   \\
						CFAverage                    & 0.029731 &    $\ \ \ \ \ \ \ \ \ \ \ \ \ \ \ h$   \\
          \end{tabular} 
         &
          \begin{tabular}{c|c|l}
           method & est.mean & group \\ \hline
						HierRCL$\chi^{2}$            & 0.127758 &    $a$  \\
						MTWL$_{idf}$                 & 0.115792 &    $a\ b$  \\
						MTWL$_{raw}$                 & 0.111198 &    $a\ b\ c$  \\
						RLUM                         & 0.107608 &    $a\ b\ c$   \\
						RCL$\chi^{2}$                & 0.104528 &    $a\ b\ c\ d$  \\
						RCLJSD                       & 0.104528 &    $a\ b\ c\ d$  \\
						ICWL$_{idf}$                 & 0.102517 &    $a\ b\ c\ d$  \\
						HierICWL$_{raw}$             & 0.096039 &    $\ \ b\ c\ d\ e$  \\
						Popescul\&Ungar              & 0.089642 &    $\ \ b\ c\ d\ e$  \\
						HierMTWL$_{raw}$             & 0.088041 &    $\ \ b\ c\ d\ e$   \\
						ICWL$_{raw}$                 & 0.085584 &    $\ \ \ \ c\ d\ e$  \\
						CFAverage                    & 0.078762 &    $\ \ \ \ \ \ d\ e$  \\
						HierRCLJSD                   & 0.070343 &    $\ \ \ \ \ \ \ \ e$   \\
						HierMTWL$_{idf}$             & 0.068618 &    $\ \ \ \ \ \ \ \ e$  \\
						HierICWL$_{idf}$             & 0.068618 &    $\ \ \ \ \ \ \ \ e$  \\
						CFLeaveOneOut                & 0.044474 &    $\ \ \ \ \ \ \ \ \ \ f$ \\ 
          \end{tabular}  \\ \hline \hline
         IFM  & Physics  \\
          (df=11980, $\widehat{V(E)}$=0.0401, $\alpha$=0.05)  & (df=10217, $\widehat{V(E)}$=0.0295, $\alpha$=0.05)     \\ \hline
          \begin{tabular}{c|c|l}
           method & est.mean & group \\ \hline
						RCLJSD                       & 0.288323 &    $a$  \\
						RCL$\chi^{2}$                & 0.288323 &    $a$  \\
						RLUM                         & 0.142031 &    $\ \ b$  \\
						HierRCLJSD                   & 0.100995 &    $\ \ \ \ c$     \\
						Popescul\&Ungar              & 0.099465 &    $\ \ \ \ c$   \\
						HierRCL$\chi^{2}$            & 0.091832 &    $\ \ \ \ c$   \\
						MTWL$_{idf}$                 & 0.084198 &    $\ \ \ \ c\ d$   \\
						MTWL$_{raw}$                 & 0.081345 &    $\ \ \ \ c\ d$  \\
						ICWL$_{idf}$                 & 0.073422 &    $\ \ \ \ c\ d$ \\
						HierMTWL$_{raw}$             & 0.072081 &    $\ \ \ \ c\ d$  \\
						ICWL$_{raw}$                 & 0.071838 &    $\ \ \ \ c\ d$  \\
						HierICWL$_{raw}$             & 0.071358 &    $\ \ \ \ c\ d$   \\
						CFLeaveOneOut                & 0.056609 &    $\ \ \ \ \ \ d\ e$  \\
						HierMTWL$_{idf}$             & 0.036630 &    $\ \ \ \ \ \ \ \ e\ f$  \\
						HierICWL$_{idf}$             & 0.036630 &    $\ \ \ \ \ \ \ \ e\ f$   \\
						CFAverage                    & 0.025167 &    $\ \ \ \ \ \ \ \ \ \ f$   \\
				  \end{tabular}
				  &
				  \begin{tabular}{c|c|l}
           method & est.mean & group \\ \hline
						RCLJSD                       & 0.129316 &    $a$  \\
						RCL$\chi^{2}$                & 0.128909 &    $a$  \\
						RLUM                         & 0.121768 &    $a\ b$  \\
						HierRCL$\chi^{2}$            & 0.119685 &    $a\ b$   \\
						MTWL$_{idf}$                 & 0.113828 &    $a\ b\ c$ \\
						MTWL$_{raw}$                 & 0.101872 &    $\ \ b\ c\ d$  \\
						ICWL$_{idf}$                 & 0.099402 &    $\ \ b\ c\ d$  \\
						Popescul\&Ungar              & 0.098935 &    $\ \ b\ c\ d$   \\
						ICWL$_{raw}$                 & 0.094610 &    $\ \ b\ c\ d$  \\
						HierICWL$_{raw}$             & 0.094196 &    $\ \ b\ c\ d$   \\
						HierMTWL$_{raw}$             & 0.089644 &    $\ \ \ \ c\ d$   \\
						HierICWL$_{idf}$             & 0.079105 &    $\ \ \ \ \ \ d\ e$ \\
						HierMTWL$_{idf}$             & 0.079105 &    $\ \ \ \ \ \ d\ e$ \\
						HierRCLJSD                   & 0.062467 &    $\ \ \ \ \ \ \ \ e$\\
						CFLeaveOneOut                & 0.042974 &    $\ \ \ \ \ \ \ \ \ \ f$ \\
						CFAverage                    & 0.039063 &    $\ \ \ \ \ \ \ \ \ \ f$   \\
				  \end{tabular}
      \end{tabular} \\ \hline
 \end{tabular}}
\end{table}  


We are assuming that generic labels interpreted as search queries ($Q(L_g)$) are able to represent the hierarchical relations among the clusters. The obtained results are presented in Table \ref{resultadoMetodos:Genericos}. The methods which rank the attributes with $\chi^{2}$ and JSD, based on a cluster reference collection, have the best \textit{F measure} mean values for all text collections. However, for Computer and Physics text collections, there are methods which use global reference collection in the ``a'' $mean\ group$, as HierRCL$\chi^{2}$, RLUM and the MTWL$_{idf}$. Besides, for the Computer collection, the MTWL$_{raw}$ is also in the ``a'' $mean\ group$. These results provided some evidence that the simplest methods, which rank the attributes by $\chi^{2}$ or JSD, are sufficient to describe a topic hierarchy with some advantage over the MTWL$_{raw}$ (most frequent). On the other hand, for the biggest text collection, which results are presented in Table \ref{F:Limitations}, only $Popescul\&Ungar$, among the methods which rank the attributes with  $\chi^2$, is in the first mean group. The second mean group, composed by $MTWL_{idf}$ and $CFAverage$, also uses the global reference collection in the ranking function. Finally, $CFLeaveOneOut$  had a better performance for PubMed-Cancer than for the small text collections. In this way, there is an evidence that for a big text collection it is better to use methods which consider the global reference collection. Additionally, only the  MTWL$_{idf}$ had the same behavior in a small and in a big text collection.

Ainda, essa análise também reforça as características propostas da avaliação de generic query expressions ($Q(L_g)$), isto é, priorizando os métodos de seleção de descritores que tratam o problema considerando a hierarquia ``globalmente'' (global = olha mais de um nível)

\begin{table} \centering
 \caption{Multiple mean comparisons for Pubmed-Cancer text collection - by methods.}
\label{F:Limitations}
\resizebox{0.8\textwidth}{!} {
 \begin{tabular}{|c|} \hline
   F measure\\ \hline
   \begin{tabular}{c||c}
          Specific label queries  & Generic label queries  \\
          (df=2111398, $\widehat{V(E)}$=0.0000, $\alpha$=0.05)  & (df=2111398, $\widehat{V(E)}$=0.0001, $\alpha$=0.05)     \\ \hline
          \begin{tabular}{c|c|l}
           method                                       & est.mean      & group \\ \hline
           
Popescul\&Ungar & 0.000992   & $a$ \\
MTWL$_{idf}$ & 0.000898      & $\ b$ \\
ICWL$_{idf}$ & 0.000869      & $\ \ c$ \\
CFLeaveOneOut & 0.000852     & $\ \ c$ \\
HierMTWL$_{idf}$ & 0.000775  & $\ \ \ d$ \\
HierICWL$_{idf}$ & 0.000760  & $\ \ \ d$ \\
CFAverage & 0.000732         & $\ \ \ \ e$ \\
MTWL$_{raw}$ & 0.000687      & $\ \ \ \ \ f$ \\
ICWL$_{raw}$ & 0.000686      & $\ \ \ \ \ f$ \\
HierICWL$_{raw}$ & 0.000674  & $\ \ \ \ \ f$ \\
HierRCL$\chi^{2}$ & 0.000669 & $\ \ \ \ \ f$ \\
HierMTWL$_{raw}$ & 0.000668  & $\ \ \ \ \ f$ \\
RLUM & 0.000631              & $\ \ \ \ \ \ g$ \\
RCL$\chi^{2}$ & 0.000502     & $\ \ \ \ \ \ \ h$ \\
RCLJSD & 0.000476            & $\ \ \ \ \ \ \ \ i$ \\
HierRCLJSD & 0.000455        & $\ \ \ \ \ \ \ \ i$ \\
          \end{tabular} 
         &
          \begin{tabular}{c|c|l}
           method & est.mean & group \\ \hline
           
           	Popescul\&Ungar & 0.002377   & $a$ \\
CFAverage & 0.002257         & $\ b$ \\
MTWL$_{idf}$ & 0.002245      & $\ b$ \\
CFLeaveOneOut & 0.002146     & $\ \ c$ \\  
ICWL$_{idf}$ & 0.002007      & $\ \ \ d$ \\
HierRCL$\chi^{2}$ & 0.001874 & $\ \ \ \ e$  \\
ICWL$_{raw}$ & 0.001356      & $\ \ \ \ \ f$ \\
MTWL$_{raw}$ & 0.001327      & $\ \ \ \ \ f$ \\
HierICWL$_{raw}$ & 0.001245  & $\ \ \ \ \ \ g$ \\
HierMTWL$_{raw}$ & 0.001212  & $\ \ \ \ \ \ g$ \\
HierICWL$_{idf}$ & 0.001199  & $\ \ \ \ \ \ g$ \\
RLUM & 0.001185              & $\ \ \ \ \ \ g$ \\
HierMTWL$_{idf}$ & 0.000953  & $\ \ \ \ \ \ \ h$ \\
RCL$\chi^{2}$ & 0.000441     & $\ \ \ \ \ \ \ \ i$ \\
HierRCLJSD & 0.000412        & $\ \ \ \ \ \ \ \ i$ \\
RCLJSD & 0.000320            & $\ \ \ \ \ \ \ \ \ j$ \\ 
          \end{tabular}  \\      
      \end{tabular} \\ \hline
 \end{tabular}}
\end{table}

\subsection{Results of retrieval measure variances among the hierarchical levels}
\label{ResVarHier}


It is interesting to evaluate each cluster labeling method among the hierarchical effects aiming to understand how the level effects are in different $mean\ groups$. In section \ref{sec:glm}, there is an explanation about how to get those values.

The level effects can be observed in the graphics constructed for each text collection and cluster labeling method using the calculated values for \textit{F measures} resulting from $Q(L_s)$ and $Q(L_g)$. We focused on the methods which had the best \textit{F measures} in Table \ref{resultadoMetodos:Genericos}. These methods are being considered the most interesting, because they seem to have selected the best generic labels ($L_g$). 

\begin{figure}[!h]
   	\centering
	\includegraphics[width=0.9\textwidth]{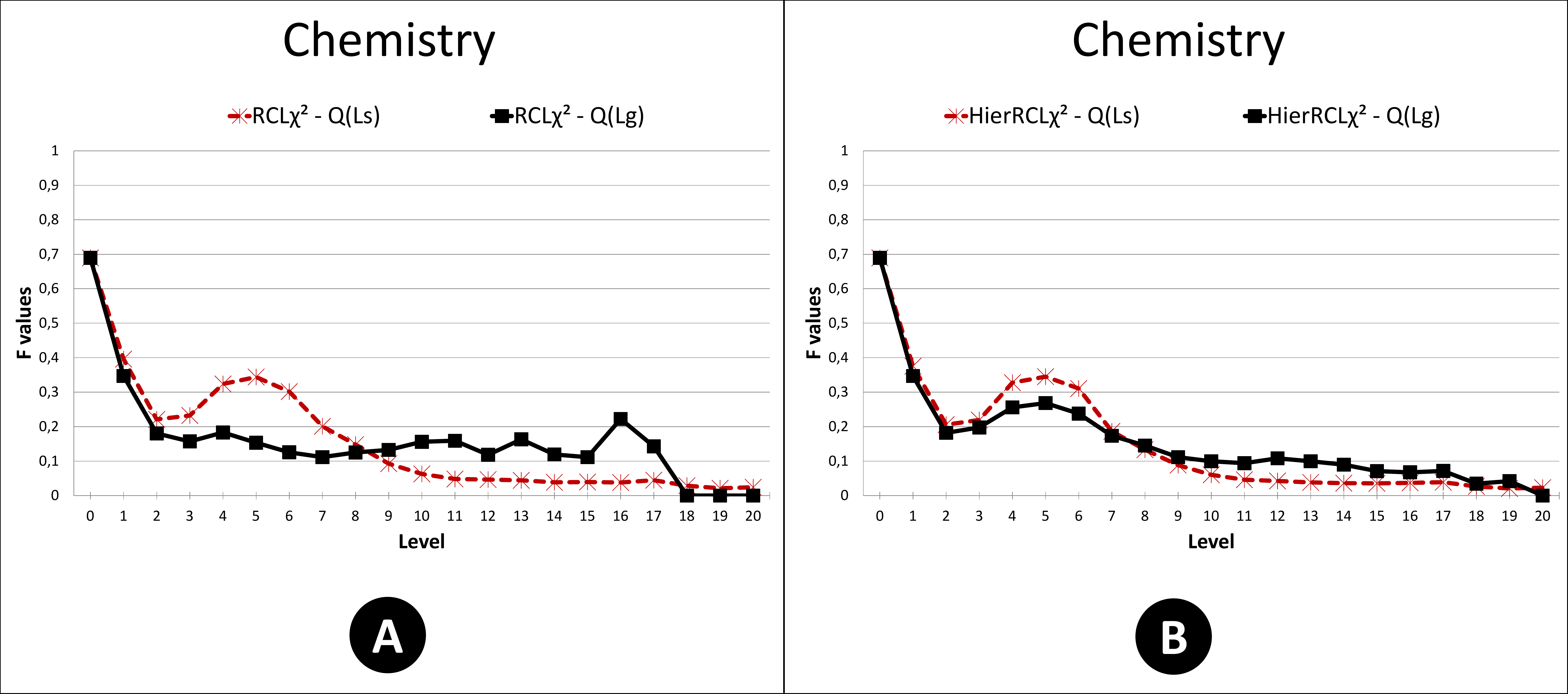}
	\caption{Chemistry collection -  $Q(L_s)$ and $Q(L_g)$ in hierarchical levels - RCL$\chi^2$ and HierRCL$\chi^2$methods.}
	\label{fig:chem1LsLg}
\end{figure}
\begin{figure}[!h]
 	\centering
	\includegraphics[width=0.9\textwidth]{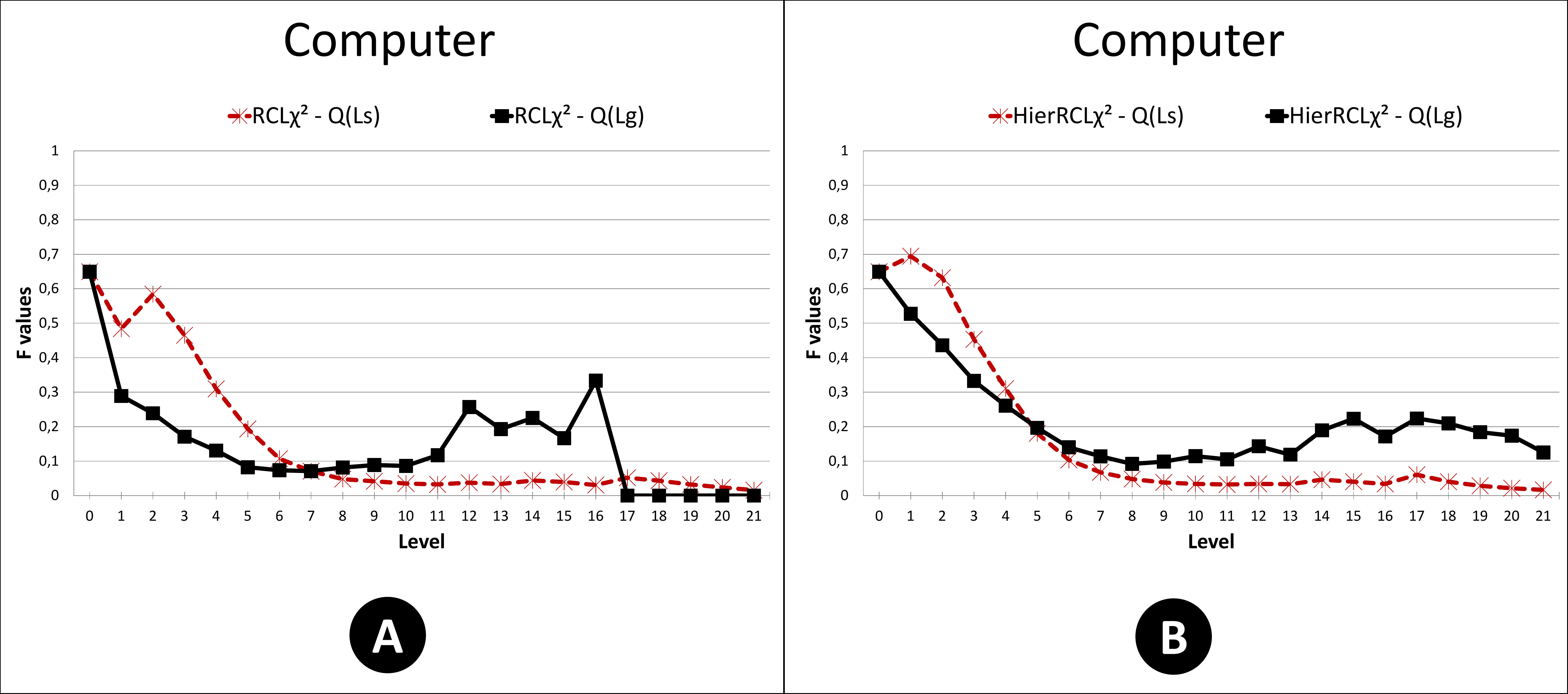}
	\caption{Computer collection  -  $Q(L_s)$ and $Q(L_g)$ in hierarchical levels - RCL$\chi^2$ and HierRCL$\chi^2$ methods.}
	\label{fig:comp1LsLg}
\end{figure}

Firstly, let us observe the results for the methods RCL$\chi^{2}$ and HierRCL$\chi^2$ in Figures \ref{fig:chem1LsLg} and \ref{fig:comp1LsLg}. For the Chemistry collection in Figure \ref{fig:chem1LsLg}, we can observe the behavior of the RCL$\chi^2$ method versus the HierRCL$\chi^2$. The behavior of RCL$\chi^2$ was better in the multiple mean comparison but, when using the ``Hier'' weight with $Q(L_s)$ or $Q(L_g)$, the \textit{F curve} has a smoother behavior and the most specific information can be captured. Consequently, the \textit{F measure} means are better when using only the cluster reference collection, but the results are improved when adding the global collection reference effects. 
 
\begin{figure}[!h]
   	\centering
	\includegraphics[width=0.9\textwidth]{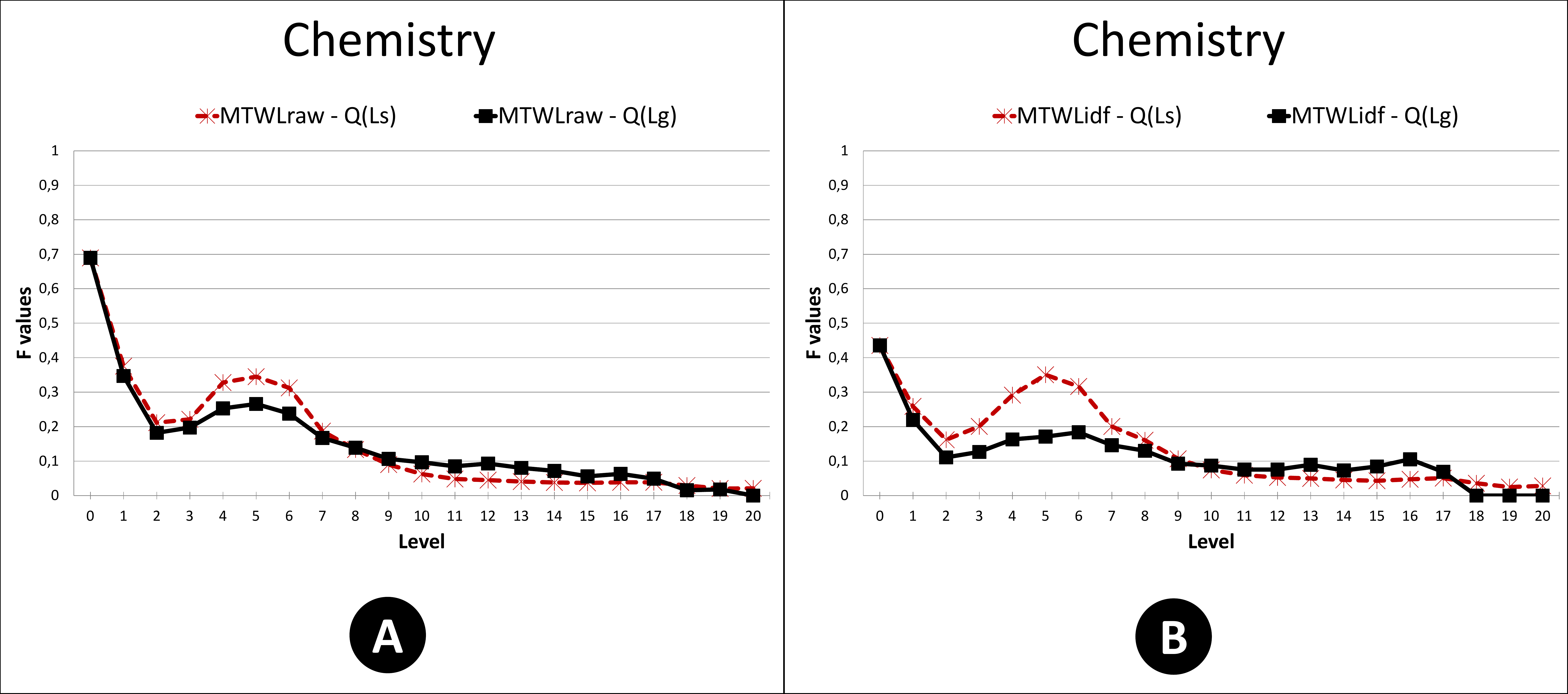}
	\caption{Chemistry collection  -  $Q(L_s)$ and $Q(L_g)$ in hierarchical levels - MTWL$raw$ and MTWL$idf$ methods.}
	\label{fig:chem2LsLg}
\end{figure}

\begin{figure}[!h]
 	\centering
	\includegraphics[width=0.9\textwidth]{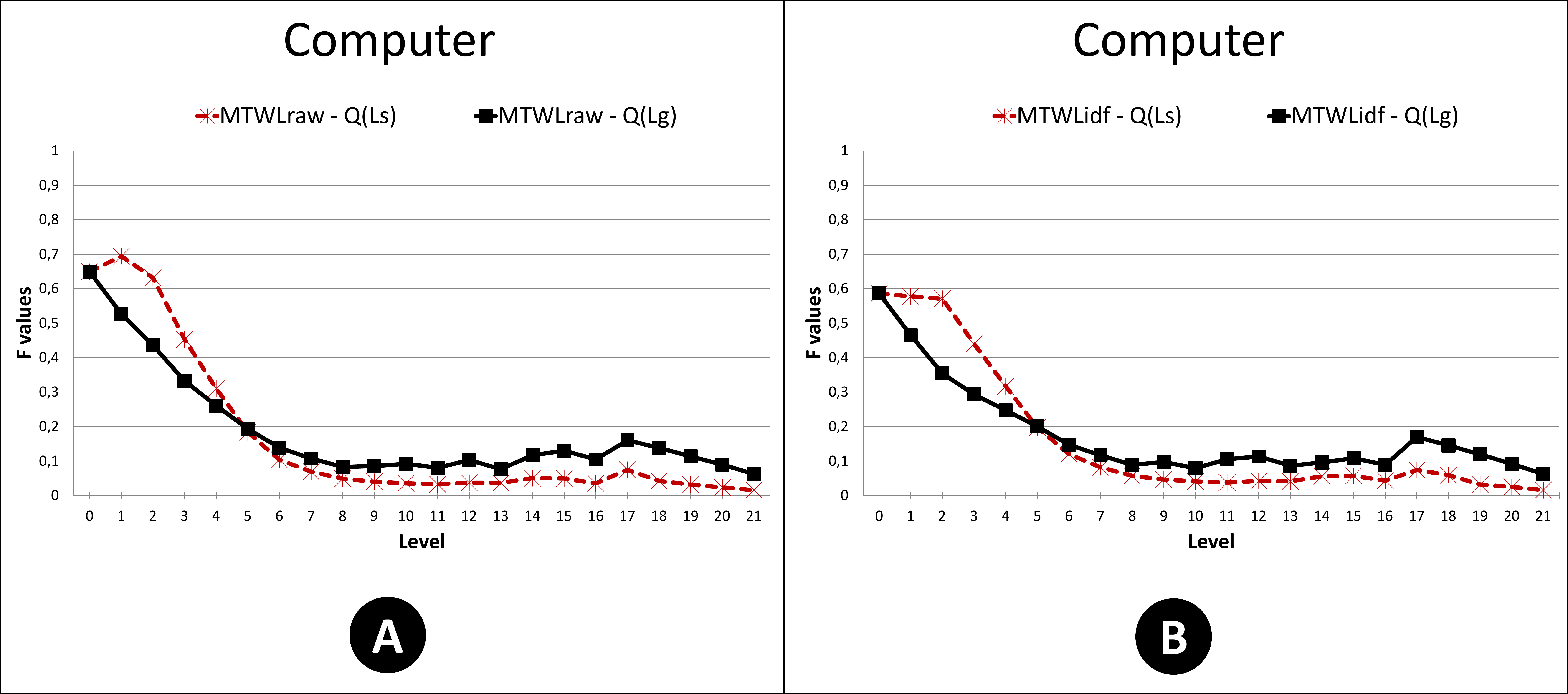}
	\caption{Computer 	collection  -  $Q(L_s)$ and $Q(L_g)$ in hierarchical levels - MTWL$raw$ and MTWL$idf$ methods.}
	\label{fig:comp2LsLg}
\end{figure}

In Figures \ref{fig:chem2LsLg} and \ref{fig:comp2LsLg}, the \textit{F values} for the methods MTWL$_{raw}$ and MTWL$_{idf}$, using $Q(L_s)$ and $Q(L_g)$, are illustrated. We can note the proximity of the \textit{F values} from the methods in the Computer collection (Figure \ref{fig:comp2LsLg} ), which has a more balanced hierarchy. Besides, we obtain better results using $Q(L_s)$ for the Chemistry collection (Figure \ref{fig:chem2LsLg}), which can be due to the greater independence among the groups. These observations offer evidence that the MTWL$_{raw}$ method is very efficient for flat clusters, but not so good at capturing the topic hierarchical relations. Nevertheless, there is a slightly better performance for the methods from the middle to the leaves of the hierarchy using $Q(L_g)$. Furthermore, in Figure \ref{fig:chem2LsLg}(A) and \ref{fig:chem2LsLg}(B), we can notice that the \textit{F curve} for $Q(L_s)$ is very similar to the correspondent curve in Figure \ref{fig:chem1LsLg}(A) and Figure \ref{fig:chem1LsLg}(B). That is, the RCL$\chi^2$ and MTWL$raw$ had a very similar behavior when using $Q(L_s)$. On the other hand, we must note that in the generic case, using $Q(L_g)$, Figure \ref{fig:chem2LsLg}(B), MTWL$idf$ had a worse performance than HierRCL$\chi^2$ (with $Q(L_g)$ in Figure \ref{fig:chem2LsLg}(B)).

\begin{figure}[!h]
 	\centering
    \includegraphics[width=0.9\textwidth]{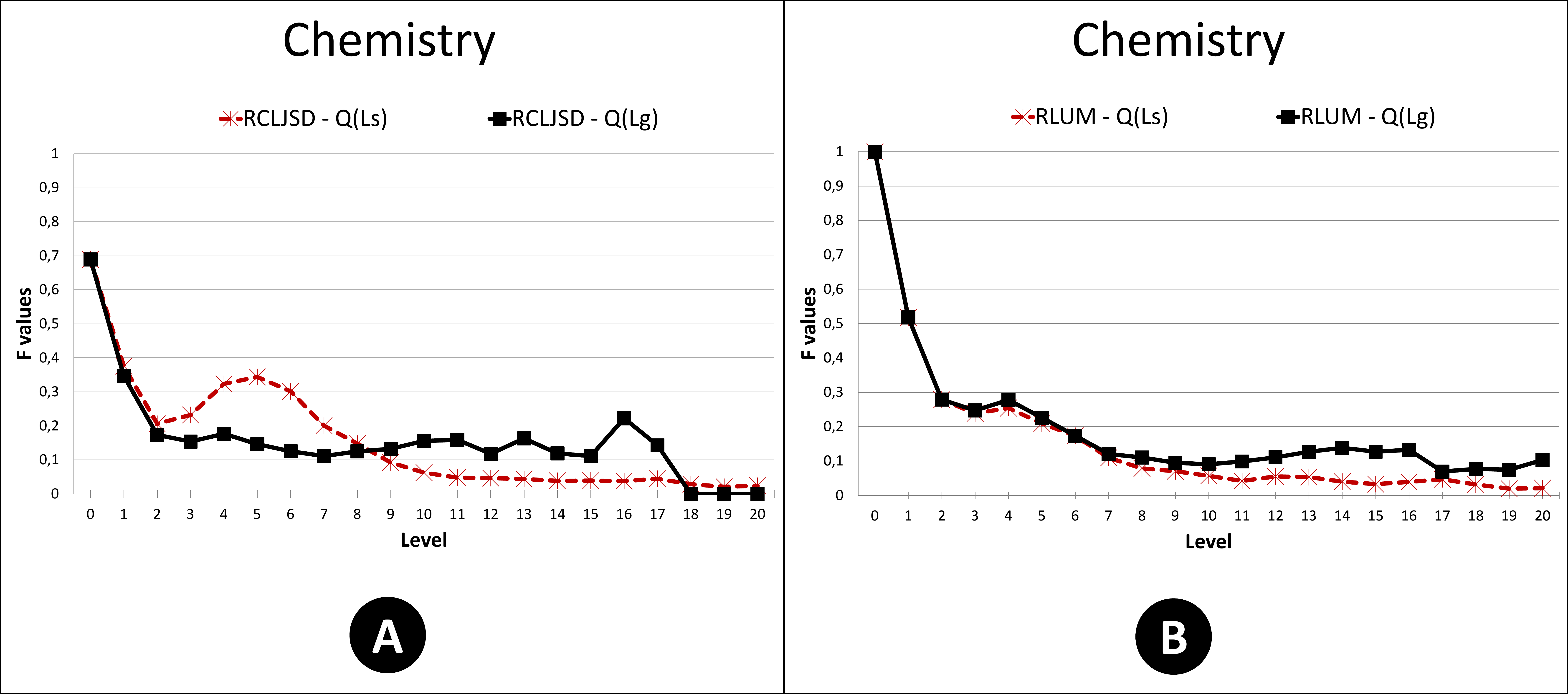}
	\caption{Chemistry collection  -  $Q(L_s)$ and $Q(L_g)$ in hierarchical levels - RCLJSD and RLUM methods.}
	\label{fig:chem3LsLg}
\end{figure}

\begin{figure}[!h]
 	\centering
	\includegraphics[width=0.9\textwidth]{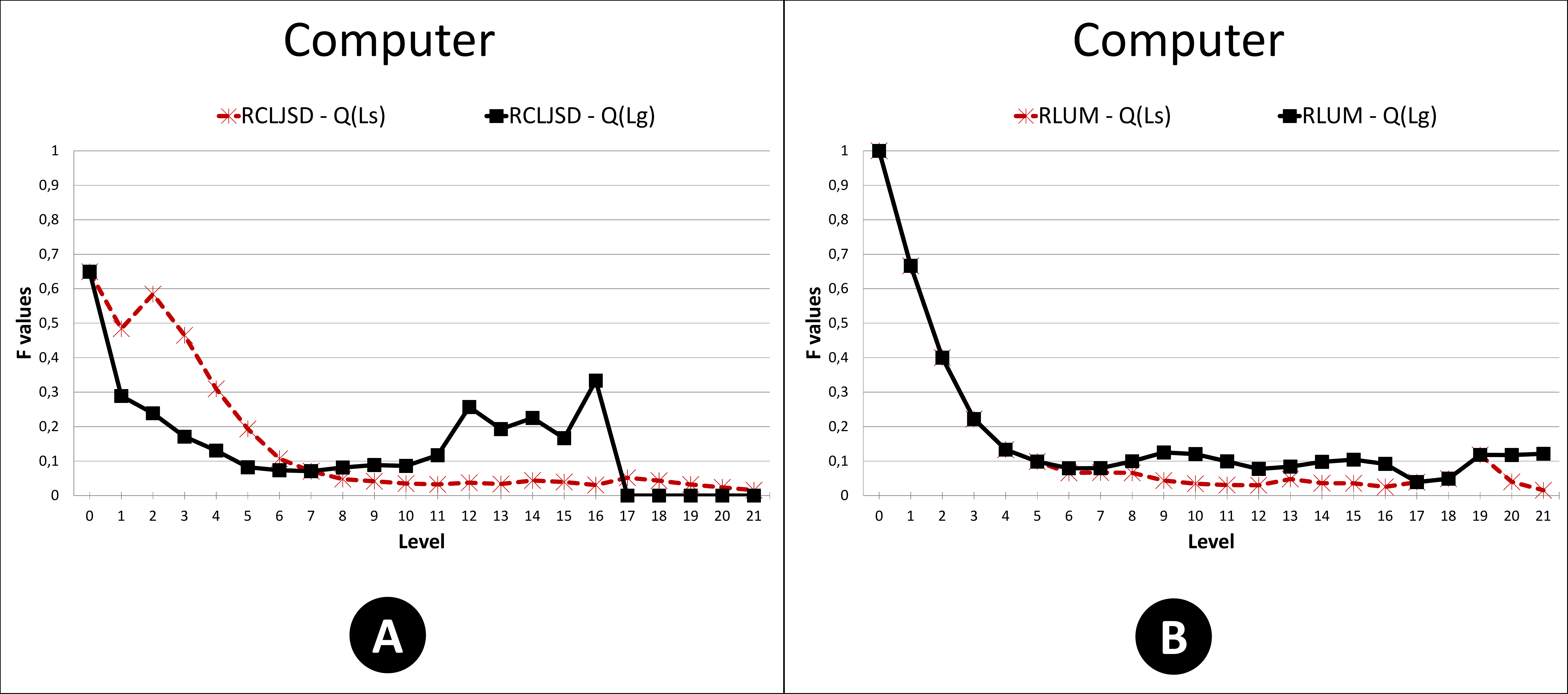}
	\caption{Computer collection -  $Q(L_s)$ and $Q(L_g)$ in hierarchical levels - RCLJSD and RLUM methods.}
	\label{fig:comp3LsLg}
\end{figure}

In Figures \ref{fig:chem3LsLg} and \ref{fig:comp3LsLg}, the \textit{F values} for the methods RCLJSD and RLUM, using $Q(L_s)$ and $Q(L_g)$, are illustrated. We must note that the behavior of RCLJSD is almost the same as that of RCL$\chi^2$ (Figures \ref{fig:chem1LsLg}(A) and \ref{fig:comp1LsLg}(A)), for both text collections. Moreover, RCLJSD and RCL$\chi^2$ fail at capturing the information close to the leaves, that is, the most specific information in the topic hierarchy, when using the generic query ($Q(L_g)$). Also, we must note that the RLUM method has the same behavior in both text collections. In comparison to the other methods, RLUM has a good performance in capturing both the most generic and the most specific information (using $Q(L_g)$). However, the RCLJSD got the best \textit{F values}. 
%

As we could observe, the level effects for different methods are in different $mean$ $groups$ - Tables \ref{resultadoNiveis:Genericos} and \ref{resultadoNiveis:Especificos}. Moreover, this difference is clear when using $Q(L_g)$. Besides, we could note that the ``Hier'' weights and the constructed methods based on the global reference collection seem to have a smooth \textit{F curve}. Thus, it is interesting to compare these methods with their versions which were based on the local reference collection or with methods constructed in a similar manner - such as Popescul\&Ungar and RLUM, for example.

\begin{figure}[!h]
 	\centering
	\includegraphics[width=0.9\textwidth]{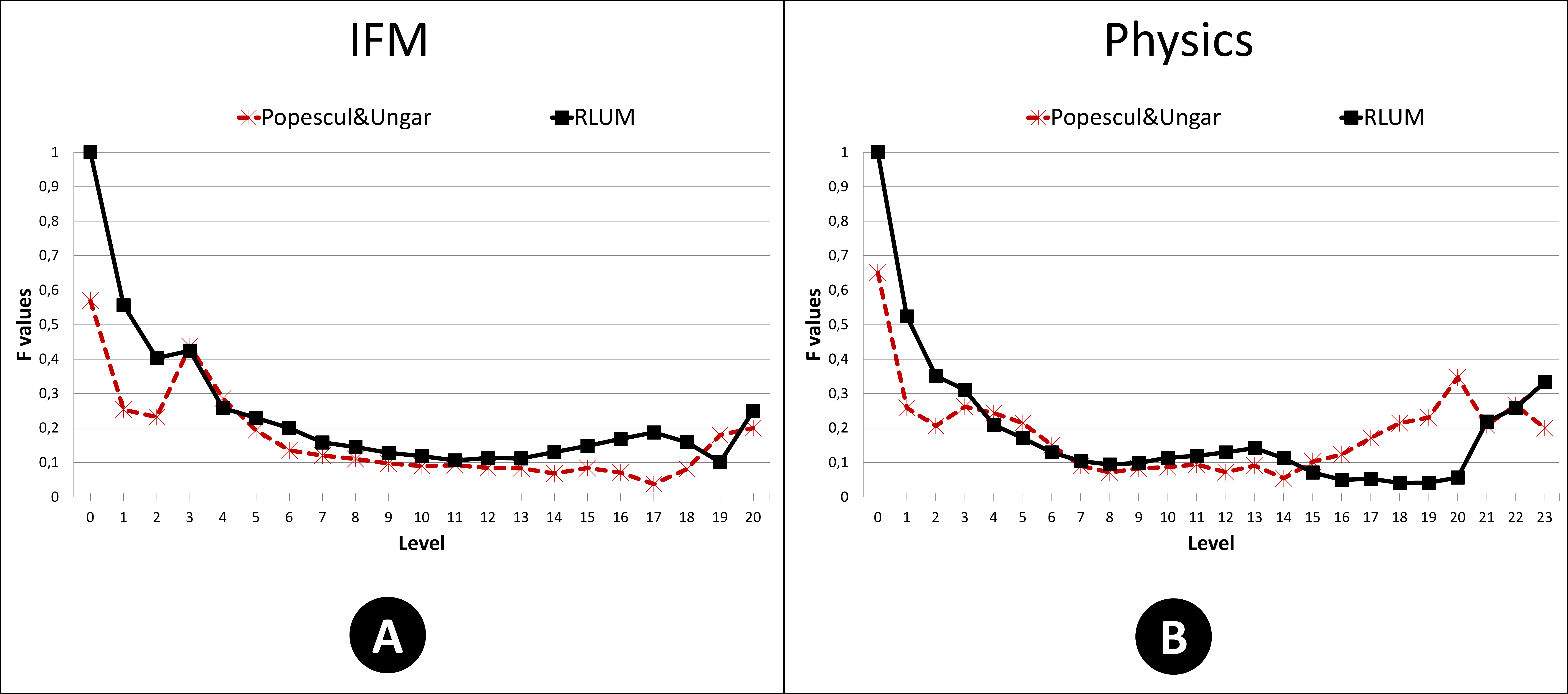}
	\caption{IFM and Physics collections -  $Q(L_g)$ - RLUM and Popescul\&Ungar methods.}
	\label{fig:LgRLUMPope}
\end{figure}

The \textit{F values} illustrated in Figure \ref{fig:LgRLUMPope} correspond to the retrieval results using $Q(L_g)$ from methods Popescul\&Ungar and RLUM, and for IFM and Physics collections. We must note that RLUM has the best values on the top and leaves of the topic hierarchies. Moreover, RLUM had some better results from the middle to the leaves of the topic hierarchy for IFM, which has an unbalanced hierarchy. On the other hand, Popescul\&Ungar had some better \textit{F values} in the middle of the topic hierarchy for the Physics collection, with a more balanced dendrogram. Generally speaking, both methods have a smooth performance and, statistically, RLUM had better results.

\begin{figure}[!h]
 	\centering
	\includegraphics[width=0.9\textwidth]{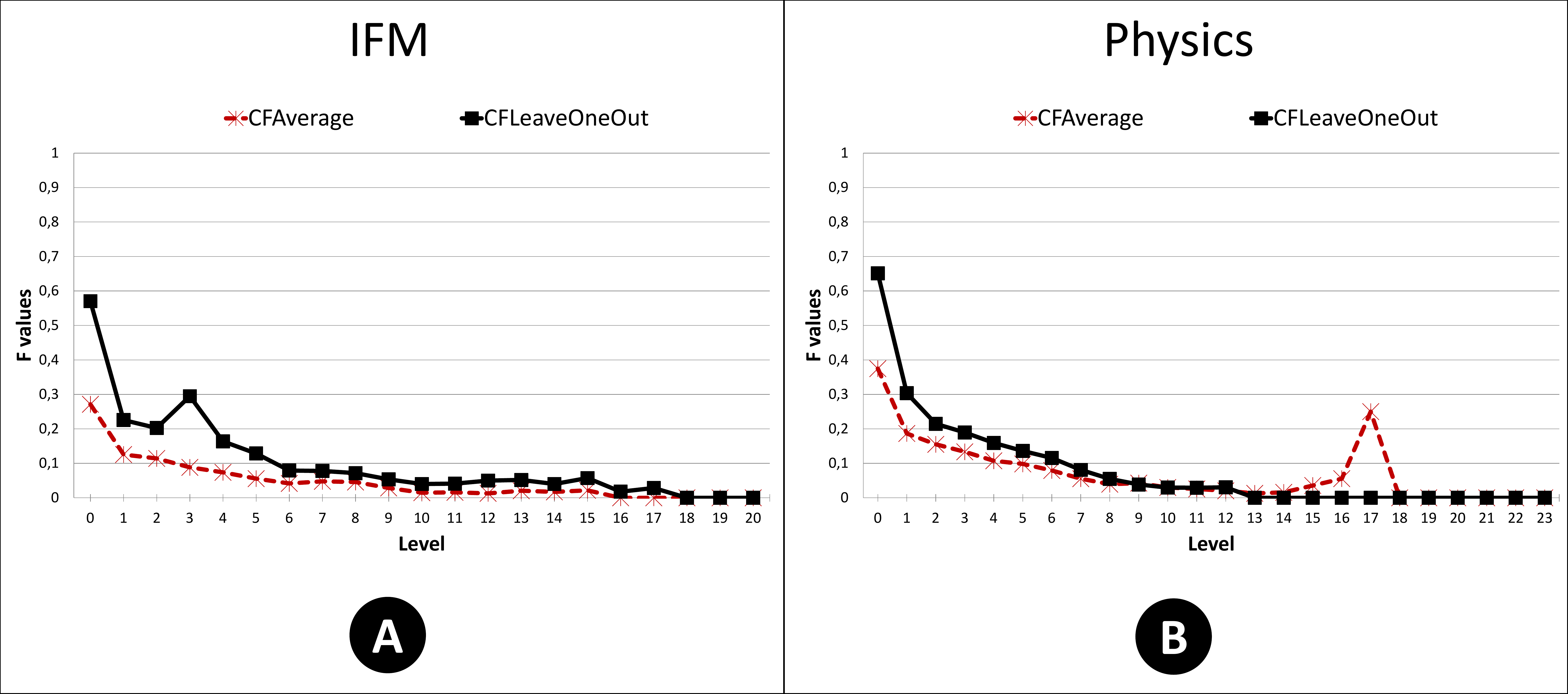}
	\caption{IFM and Physics collections -  $Q(L_g)$ - CFAverage and CFLeaveOneOut methods.}
	\label{fig:LgCFAveOut}
\end{figure}

The \textit{F values} illustrated in Figure \ref{fig:LgCFAveOut} are calculated from the retrieval results using $Q(L_g)$ from methods CFAverage and CFLeaveOneOut for IFM and Physics collections. We must note that the label sets selected by both methods seem to be good until the middle of the hierarchy. From the middle to the leaves they fail at retrieving the information or they have some isolated high values. These methods were designed for short hierarchies to improve the hyperbolic visualization, and they actually seem to be better in the high levels. However, CFLeaveOneOut has some better results than CFAverage.

\begin{figure}
 	\centering
	\includegraphics[width=0.9\textwidth]{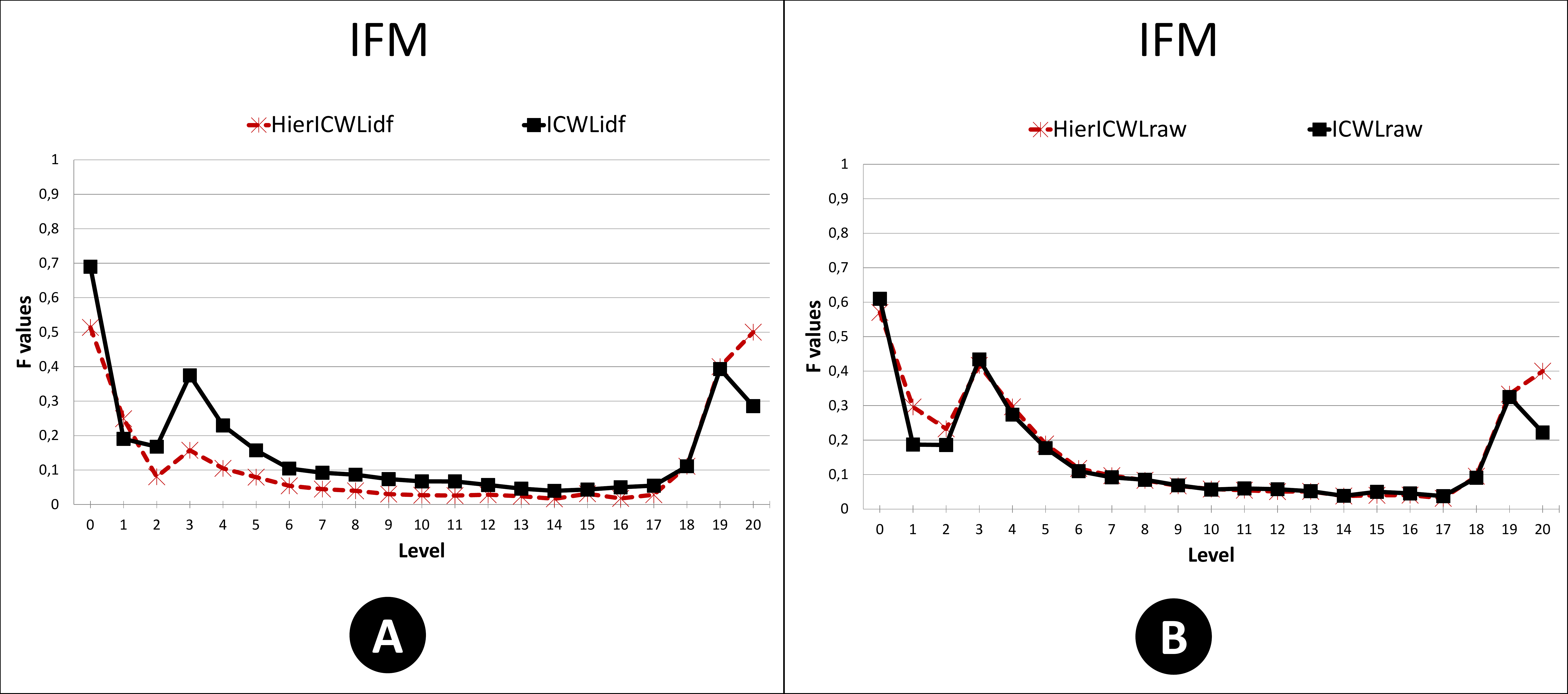}
	\caption{IFM collection -  $Q(L_g)$ -  HierICWL$_{idf}$, ICWL$_{idf}$, HierICWL$_{raw}$ and ICWL$_{raw}$ methods.}
	\label{fig:ifm2Level}
\end{figure}

The \textit{F values} in Figure \ref{fig:ifm2Level} were obtained from the retrieval results using $Q(L_g)$ from methods HierICWL$_{idf}$, ICWL$_{idf}$, HierICWL$_{raw}$ and ICWL$_{raw}$ for the IFM collection. We must note that the ``Hier'' weight was not significant in both graphics, and that the same effect was observed in the other text collections. In fact, the differences are in the root and in the leaf nodes. That is, there is no observed difference when using the ``Hier'' weights with ICWL$_{idf}$ or ICWL$_{raw}$ in our experiments. 

\begin{figure}[!h]
 	\centering
	\includegraphics[width=0.9\textwidth]{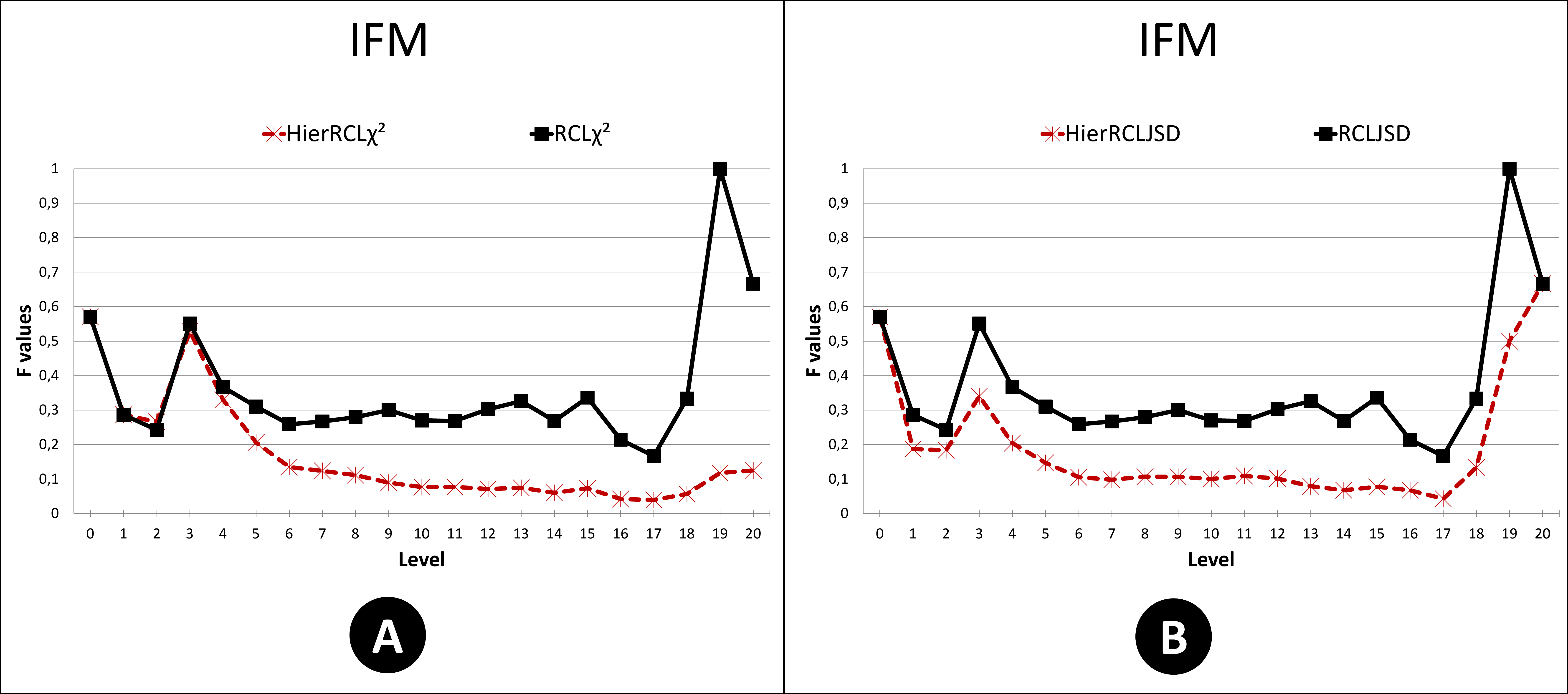}
	\caption{IFM collection -  $Q(L_g)$ - HierRCL$\chi^2$, RCL$\chi^2$, HierRCLJSD and RCLJSD methods.}
	\label{fig:ifm4Level}
\end{figure}

\begin{figure}[!h]
 	\centering
	\includegraphics[width=0.9\textwidth]{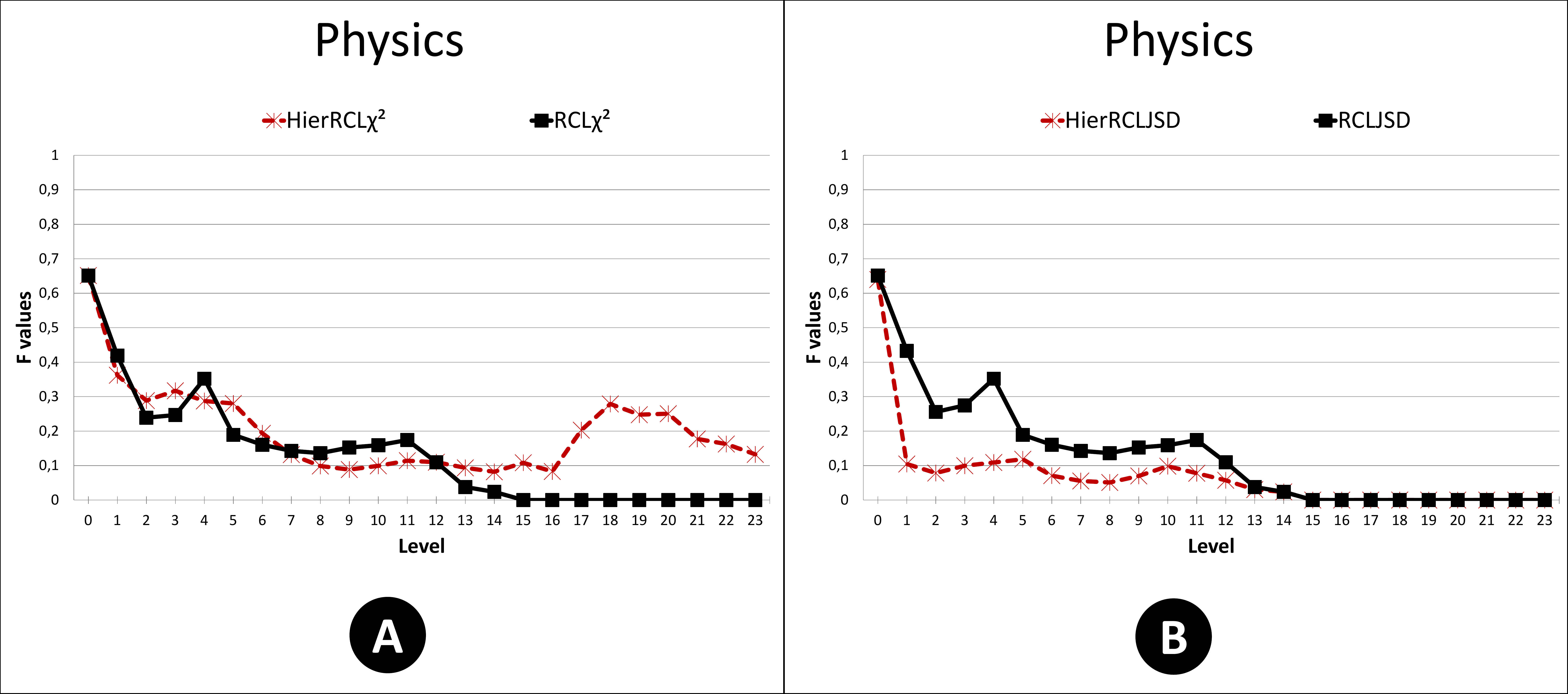}
	\caption{Physics collection -  $Q(L_g)$ - HierRCL$\chi^2$, RCL$\chi^2$, HierRCLJSD and RCLJSD methods.}
	\label{fig:phy4Level}
\end{figure}

In Figures \ref{fig:ifm4Level} and \ref{fig:phy4Level} we can observe the \textit{F values} for the retrieval results using $Q(L_g)$ from methods HierRCL$\chi^2$, RCL$\chi^2$, HierRCLJSD and RCLJSD for IFM and Physics collections. These methods presented different behaviors for balanced and unbalanced hierarchies. On the other hand, the RCLJSD and RCL$\chi^2$ had very similar values for the IFM and Physics collections. Furthermore, it is interesting to note that the ``Hier'' weight was good for the RCL$\chi^2$ for the Physics collection, with a better performance close to the leaves. However, this behavior is not true in the other collections or methods. That is, the ``Hier'' weight, when applied in a balanced hierarchy, tends to be better at capturing the more specific information as well as the more general information. This could be observed in Figure \ref{fig:comp1LsLg} too, for the Computer collection. 

\begin{figure}
 	\centering
	\includegraphics[width=0.9\textwidth]{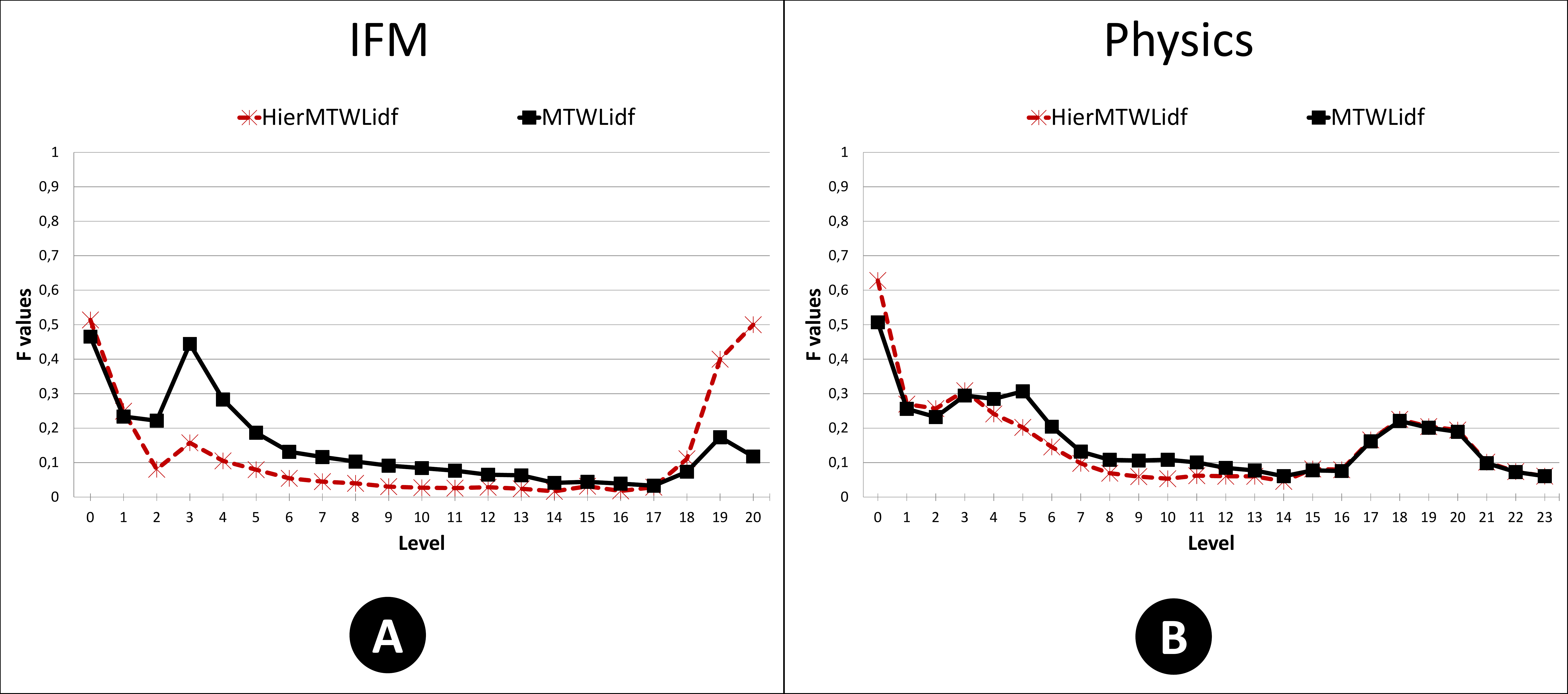}
	\caption{IFM and Physics collections -  $Q(L_g)$ - HierMTWL$_{idf}$ and MTWL$_{idf}$ methods.}
	\label{fig:LgIFMPhy}
\end{figure}

In Figure \ref{fig:LgIFMPhy} we can observe the \textit{F values} for the retrieval results using $Q(L_g)$ from methods HierMTWL$_{idf}$ and MTWL$_{idf}$ for Physics and IFM collections. The \textit{F values} in the balanced hierarchy, Physics, were not so influenced by the ``Hier'' weight. However, for the IFM collection, which has an unbalanced hierarchy, the behavior of the values influenced by the ``Hier'' weight is clear close to the leaves. For all text collections in this experiment, MTWL$_{idf}$ had a better performance than the HierMTWL$_{idf}$. 

Again, we observed that the behavior of the methods MTWL$_{raw}$ and HierMTWL$_{raw}$ was very similar. Thus, the MTWL$_{raw}$ method is very stable. 

\begin{figure}
\centering
\includegraphics[width=0.9\textwidth]{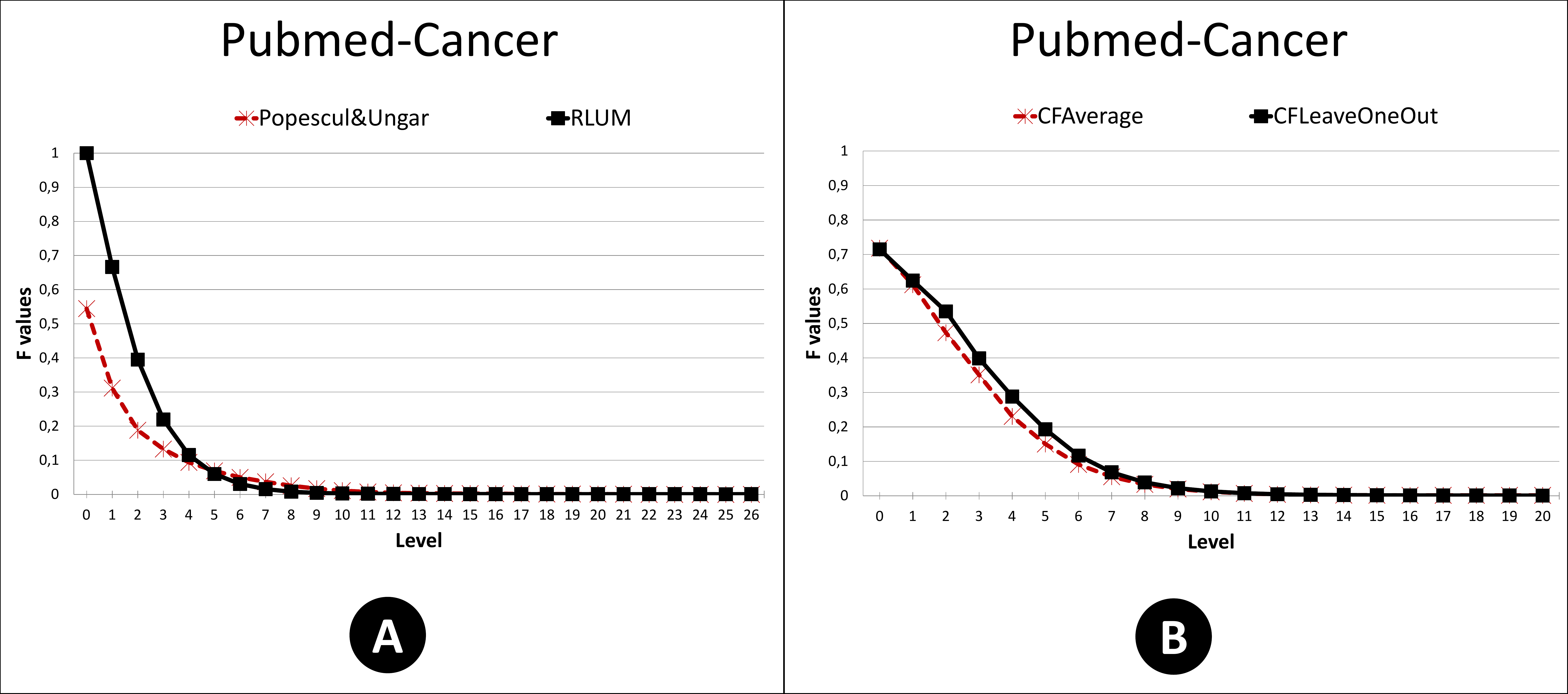}
\caption{$Q(L_g)$ - RLUM, Popescul\&Ungar, CFAverage and CFLeaveOneOut methods.}
\label{fig:LgCFAveOutRlumPope}
\end{figure}

\begin{figure}
\centering
\includegraphics[width=0.9\textwidth]{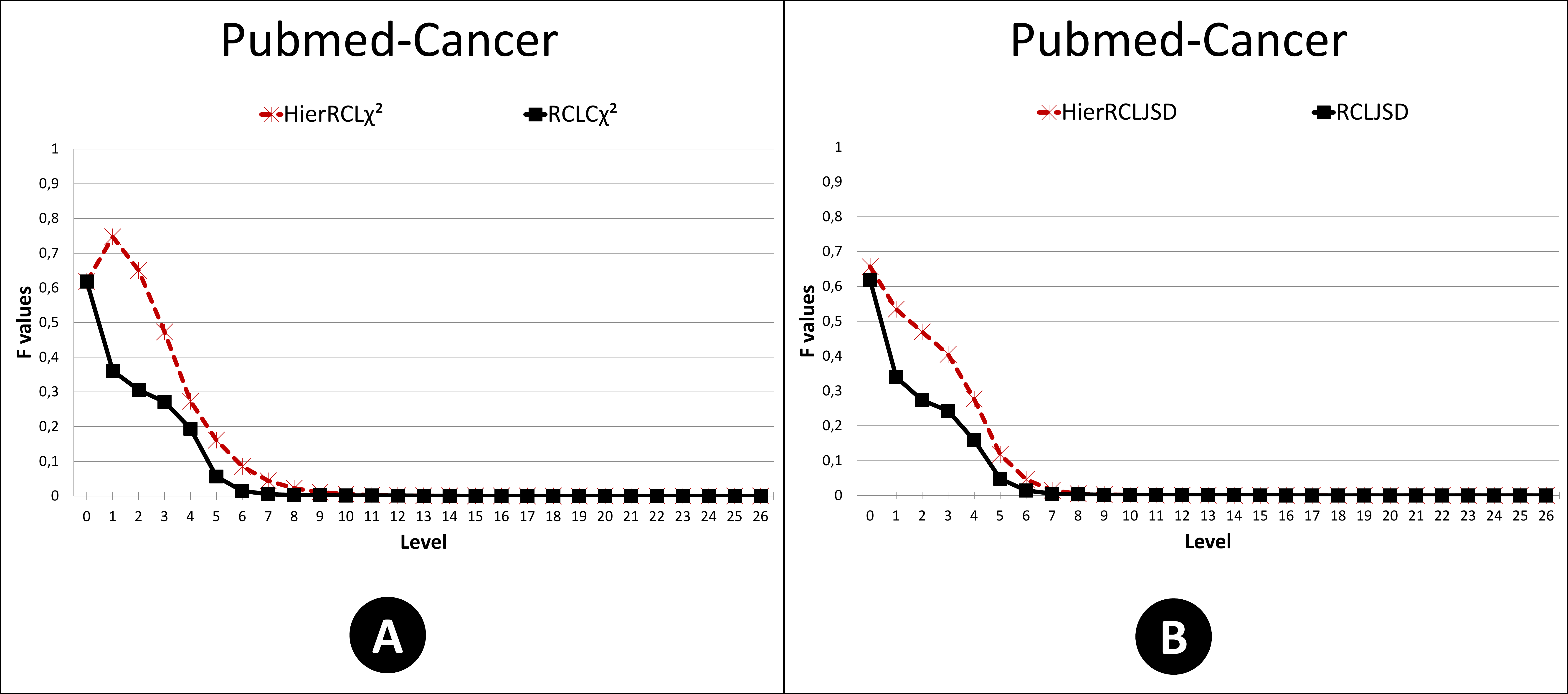}
\caption{$Q(L_g)$ - RCL$\chi^{2}$, HierRCL$\chi^{2}$, RCLJSD and HierRCLJSD methods.}
\label{fig:LgChiJSD}
\end{figure}

\begin{figure}
\centering
\includegraphics[width=0.9\textwidth]{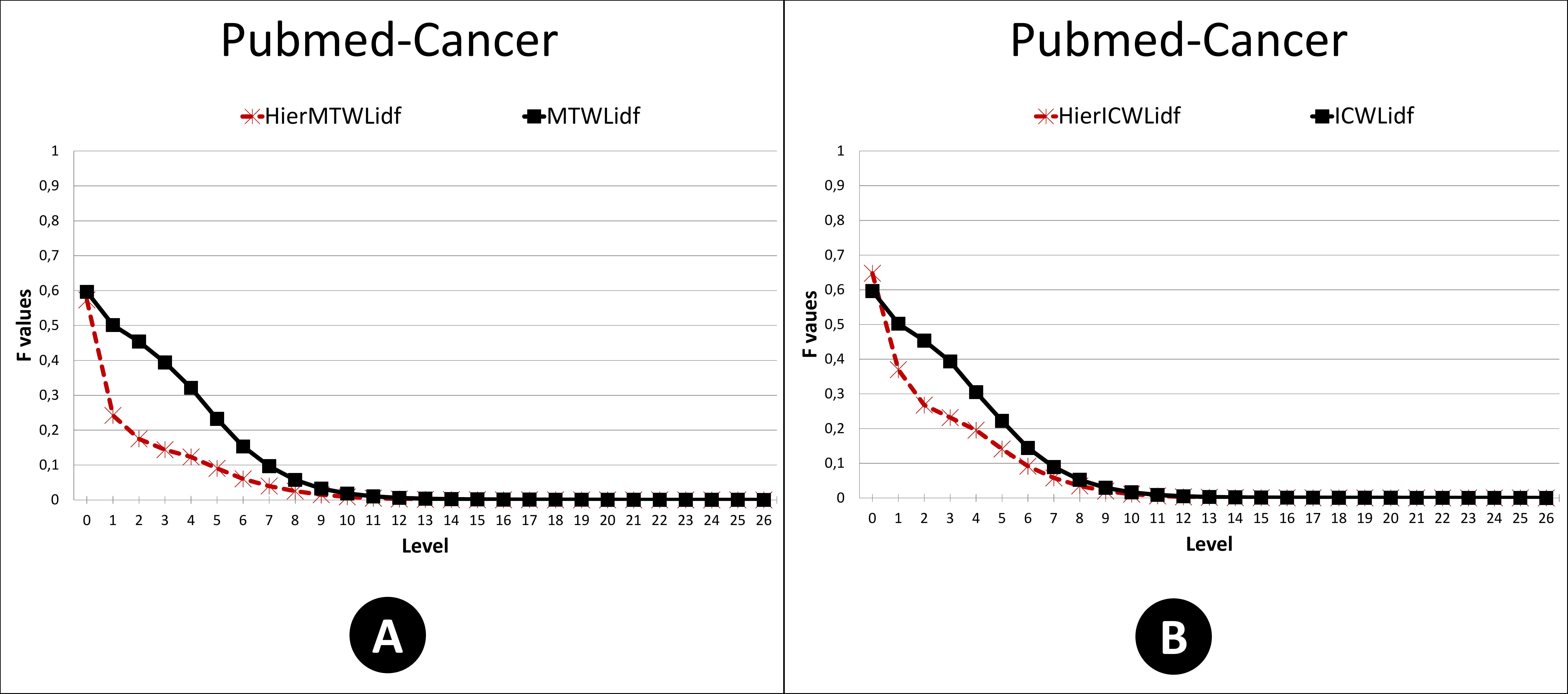}
\caption{$Q(L_g)$ - MTWL$_{idf}$, HierMTWL$_{idf}$, ICWL$_{idf}$ and HierICWL$_{idf}$ methods.}
\label{fig:LgICWL}
\end{figure}

\begin{figure}
\centering
\includegraphics[width=0.9\textwidth]{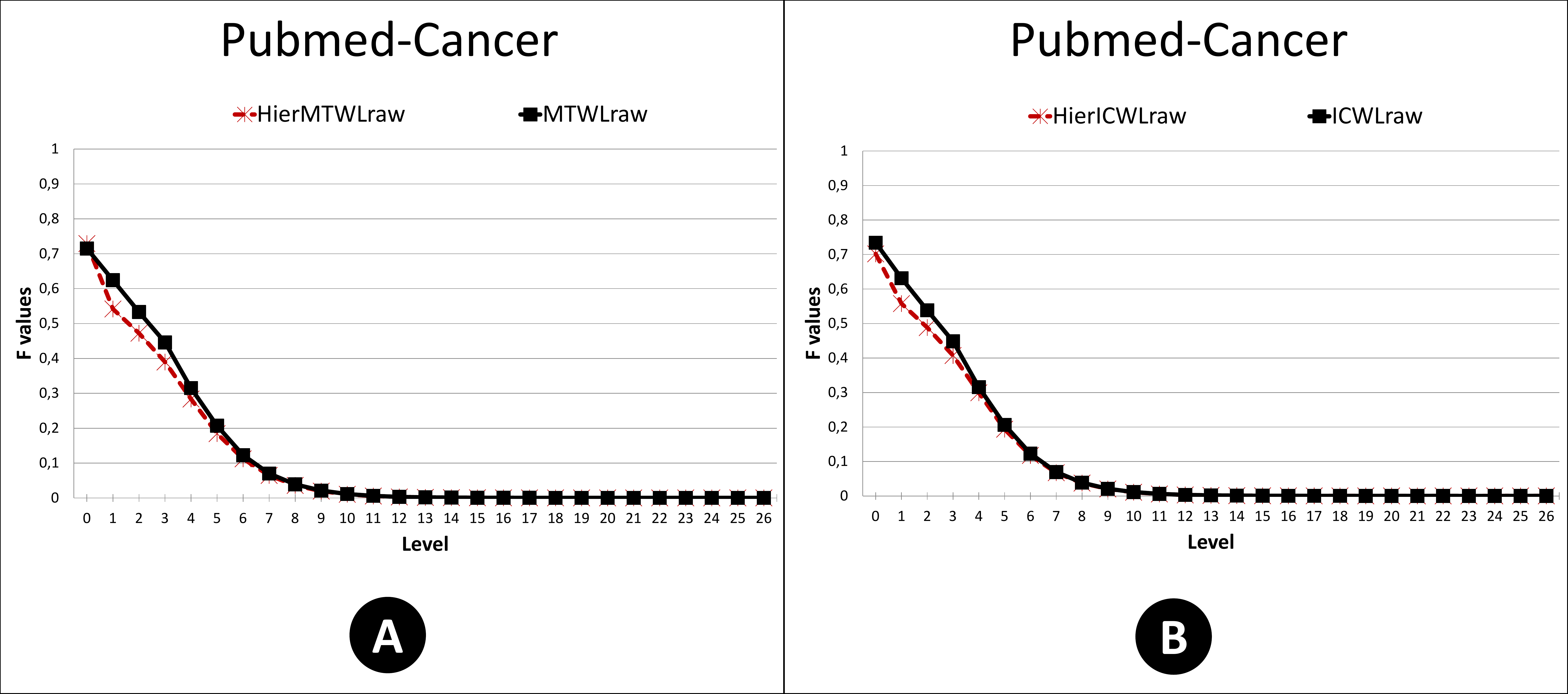}
\caption{$Q(L_g)$ - MTWL$_{raw}$, HierMTWL$_{raw}$, ICWL$_{raw}$ and HierICWL$_{raw}$ methods.}
\label{fig:LgMTWL}
\end{figure}

Finally, Figures \ref{fig:LgCFAveOutRlumPope}, \ref{fig:LgChiJSD}, \ref{fig:LgICWL} and \ref{fig:LgMTWL} present an overview of results among hierarchical levels grouping methods with similar weight schemes for PubMed-Cancer text collection. In Figure \ref{fig:LgCFAveOutRlumPope}(A), we can observe that RLUM presents a faster decreasing in the \textit{F values} than   Popescul\&Ungar. Besides, the CFAverage and CFLeaveOneOut in Figure \ref{fig:LgCFAveOutRlumPope}(B) did not present more differences than in the small text collections - for example, in Figure \ref{fig:LgCFAveOut}(B). Although the \textit{F value} behaviors to Popescul\&Ungar and CFLeaveOneOut had some improvement in Table \ref{F:Limitations}. In Figure \ref{fig:LgChiJSD}, there were some little differences comparing to the small text collections. The RCLJSD or RCL$\chi^2$ with ``Hier'' weight had some advantages in the higher levels of the hierarchy. Although, in the small collections, the RCLJSD and RCL$\chi^2$ ranking for the label selections had better results than calculated with the ``Hier'' weight. Figure \ref{fig:LgICWL} presents the results for ICWL or MTWL schemes with ``idf'' and ``Hier'' weights. For these weights the behavior is very similar to that in the smaller text collections, although the curves are smoother. In Figure \ref{fig:LgMTWL}, without the ``idf'' weights from the methods presented in Figure \ref{fig:LgICWL}, we can observe that the cluster labeling methods kept the behavior presented in the small text collections. Additionally, in Figure \ref{fig:LgICWL}, we can observe that the ``Hier'' weight did not make a difference.         

\subsection{Topic coherence evaluation}
\label{sec:ResTopCoe}

In this section, we present the results obtained for the observed coherence of the topics. Each topic corresponds to a label set of a node ($L(n_i)$) which was obtained from one cluster labeling method. The observed coherence values were calculated from the normalized pointwise mutual information (NPMI) among the attribute pairs in a label set against the combination of the word pairs in a reference corpus - as explained in section \ref{sec:topCoe}. In Table \ref{resultadoMetodos:ObservedCoherence}, we can observe the upper quartile and maximum values of the observed coherence. 

\begin{table} \centering
 \caption{Comparison of methods order by upper quartile value of Coherence.}
\label{resultadoMetodos:ObservedCoherence}
\resizebox{0.8\textwidth}{!} {
 \begin{tabular}{|c|} \hline
   Observed Coherence\\ \hline
   \begin{tabular}{c||c}
          Chemistry  & Computer  \\ \hline
          \begin{tabular}{c|c|c}
           method & upper quartile & maximum value \\ \hline
HierICWL$_{raw}$ & 0.08  & 0.29 \\
HierMTWL$_{raw}$ & 0.08  & 0.29 \\
HierICWL$_{idf}$ & 0.07  & 0.32 \\
HierMTWL$_{idf}$ & 0.07  & 0.32 \\  \hline
MTWL$_{raw}$ & 0.07  & 0.26 \\
ICWL$_{raw}$ & 0.07  & 0.25 \\
CFAverage & 0.07  & 0.22 \\
HierRCL$\chi^{2}$ & 0.06  & 0.23 \\
ICWL$_{idf}$ & 0.06  & 0.18 \\
MTWL$_{idf}$ & 0.06  & 0.17 \\ \hline
RLUM  & 0.05  & 0.22 \\
HierRCLJSD & 0.05  & 0.20 \\
RCL$\chi^{2}$ & 0.05  & 0.19 \\
RCLJSD & 0.05  & 0.19 \\
Popescul\&Ungar   & 0.04  & 0.22 \\
CFLeaveOneOut & 0.04  & 0.19 \\
          \end{tabular} 
         &
          \begin{tabular}{c|c|c}
           method & upper quartile & maximum value \\ \hline
HierICWL$_{idf}$ & 0.05  & 0.20 \\
HierICWL$_{raw}$ & 0.05  & 0.20 \\
HierMTWL$_{idf}$ & 0.05  & 0.20 \\
HierMTWL$_{raw}$ & 0.05  & 0.18 \\  \hline
ICWL$_{idf}$ & 0.04  & 0.21 \\
MTWL$_{idf}$ & 0.04  & 0.21 \\
ICWL$_{raw}$ & 0.04  & 0.17 \\
HierRCL$\chi^{2}$ & 0.04  & 0.16 \\
CFAverage & 0.04  & 0.15 \\
MTWL$_{raw}$ & 0.04  & 0.15 \\  \hline
RLUM  & 0.03  & 0.28 \\
Popescul\&Ungar   & 0.03  & 0.18 \\
RCL$\chi^{2}$ & 0.03  & 0.16 \\
RCLJSD & 0.03  & 0.16 \\
HierRCLJSD & 0.03  & 0.12 \\
CFLeaveOneOut & 0.02  & 0.15 \\
          \end{tabular}  \\ \hline \hline
         IFM  & Physics  \\ \hline
          \begin{tabular}{c|c|c}
           method & upper quartile & maximum value \\ \hline
HierICWL$_{raw}$ & 0.03  & 0.13 \\
HierMTWL$_{raw}$ & 0.03  & 0.13 \\    \hline
CFAverage & 0.02  & 0.11 \\
HierICWL$_{idf}$ & 0.02  & 0.14 \\
HierMTWL$_{idf}$ & 0.02  & 0.14 \\
HierRCL$\chi^{2}$ & 0.02  & 0.2 \\
HierRCLJSD & 0.02  & 0.10 \\
ICWL$_{idf}$ & 0.02  & 0.09 \\
ICWL$_{raw}$ & 0.02  & 0.11 \\
MTWL$_{idf}$ & 0.02  & 0.09 \\
MTWL$_{raw}$ & 0.02  & 0.09 \\
RCL$\chi^{2}$ & 0.02  & 0.27 \\
RCLJSD & 0.02  & 0.27 \\             \hline
Popescul\&Ungar   & 0.015 & 0.11 \\
CFLeaveOneOut & 0.01  & 0.11 \\
RLUM  & 0.01  & 0.27 \\
 \end{tabular}
 &
 \begin{tabular}{c|c|c}
           method & upper quartile & maximum value \\ \hline
HierICWL$_{idf}$ & 0.09  & 0.25 \\
HierICWL$_{raw}$ & 0.09  & 0.26 \\
HierMTWL$_{idf}$ & 0.09  & 0.25 \\
HierMTWL$_{raw}$ & 0.08  & 0.24 \\   \hline
ICWL$_{raw}$ & 0.07  & 0.28 \\
MTWL$_{raw}$ & 0.07  & 0.28 \\
CFAverage & 0.06  & 0.28 \\
HierRCL$\chi^{2}$ & 0.06  & 0.23 \\
ICWL$_{idf}$ & 0.06  & 0.27 \\
MTWL$_{idf}$ & 0.06  & 0.27 \\   \hline
HierRCLJSD & 0.05  & 0.23 \\
Popescul\&Ungar   & 0.05  & 0.22 \\
CFLeaveOneOut & 0.04  & 0.28 \\
RCL$\chi^{2}$ & 0.04  & 0.21 \\
RCLJSD & 0.04  & 0.21 \\
RLUM  & 0.04  & 0.28 \\
 \end{tabular}
      \end{tabular} \\ \hline
 \end{tabular}}
\end{table}

First of all, let us take a look into the results for the collections Computer, Physics and Chemistry. In Table \ref{resultadoMetodos:ObservedCoherence}, we can observed that that the cluster labeling methods HierMTWL$_{idf}$, HierMTWL$_{raw}$, HierICWL$_{raw}$ and HierICWL$_{idf}$ had the best observed coherence values. Apparently, the ``Hier'' weights were able to reflect some hierarchical context information, which helped the selection of more coherent labels. On a second group of observed coherence values, we find the methods based on frequency and $\chi^2$ rankings, that is MTWL$_{raw}$, ICWL$_{raw}$, ICWL$_{idf}$ and MTWL$_{idf}$ as well as the CFAverage and the HierRCL$\chi^{2}$. The behavior of HierRCL$\chi^2$ can be an evidence that the $\chi^{2}$ rankings were stronger than the ``Hier'' weights. We must highlight that the major part of the maximum values for the Physics collection is in this group. The third group also has a ``Hier'' weighted method (HierRCLJSD), some methods based on JSD and $\chi^{2}$ rankings, as well as RLUM, Popescul\&Ungar and CFLeaveOneOut. Thus, there is evidence that the ``Hier'' weight did not bring a significant difference to the observed coherence values of RCLJSD. RLUM and Popescul\&Ungar are cluster labeling methods which result in a significant difference among the specific and generic topic labels. In this case, the labels have non-repeated and fewer attributes than the other cluster labeling methods, which could be the cause of low coherence values. On the other hand, RLUM had the best maximum coherence value for Computer and Physics. Another important observation is that the CFAverage had better coherence values than CFLeaveOneOut for all text collections.

The IFM text collection has an unbalanced dendrogram and had the worst topic coherence values for all methods in comparison to the other text collections. Despite those results, HierMTWL$_{raw}$ and HierICWL$_{raw}$ produced the best topic coherence values. The largest maximum coherence values were produced by RCL$\chi^{2}$, RCLJSD and RLUM. Still, the obtained values are not conclusive for this text collection. 

\begin{table} \centering
 \caption{Comparison of Observed Coherence values for Pubmed-Cancer text collection order by upper quartile value.}
\label{ObservedCoherence:Limitations}
\resizebox{0.5\textwidth}{!} {
 \begin{tabular}{|c|} \hline
   Observed Coherence\\ \hline
          \begin{tabular}{c|c|c}
           method & upper quartile & maximum value \\ \hline
           
           	HierICWL$_{idf}$ & 0,12  & 0,41 \\
						HierMTWL$_{idf}$ & 0,12  & 0,41 \\
						HierICWL$_{raw}$ & 0,12  & 0,42 \\
						HierMTWL$_{raw}$ & 0,12  & 0,40 \\ \hline
						MTWL$_{raw}$ & 0,11  & 0,40 \\
						ICWL$_{raw}$ & 0,11  & 0,40 \\
						HierRCL$\chi^{2}$ & 0,10   & 0,40 \\
						MTWL$_{idf}$ & 0,10   & 0,38 \\
						RLUM  & 0,09  & 0,80 \\
						ICWL$_{idf}$ & 0,09  & 0,41 \\ \hline
						CFAverage & 0,08  & 0,35 \\
						HierRCLJSD & 0,08  & 0,37 \\
						Popescul\&Ungar & 0,07  & 0,81 \\
						CFLeaveOneOut & 0,07  & 0,42 \\
						RCL$\chi^{2}$ & 0,07  & 0,34 \\
						RCLJSD & 0,07  & 0,34 \\
          \end{tabular}  \\ \hline
 \end{tabular}}
\end{table}

Finally, Table \ref{ObservedCoherence:Limitations} presents the upper quartile and maximum values of observed coherence for the PubMed-Cancer text collection. Almost all of the method observed coherence values had the same behavior as obtained before, except for RLUM and Popescul\&Ungar which had a little improvement.  Although, RLUM and Popescul\&Ungar methods had the maximum coherence values in all experiments.

Due to the fact that the observed coherence was equivalent to a subjective evaluation \cite{newmanTopics,lau2014machine}, it is possible to believe that the HierMTWL$_{idf}$, HierMTWL$_{raw}$, HierICWL$_{raw}$ and HierICWL$_{idf}$ cluster labeling methods had the best performance regarding the coherence criterion.

\section{Conclusion}

Although there is a significant number of solutions for the production of good hierarchical document clustering and for the extraction of topics from clusters, there are few solutions for cluster labeling which consider the hierarchical relations among the topics. In particular, there is a lack of experimental comparisons among them in the literature. Thus, we propose an evaluation methodology that provides a standard pattern to present and compare label selection methods for hierarchical document clusters. Furthermore, the methodology encompasses two types of objective evaluations. The first evaluation refers to the model ability to reflect the concept of term propagation in a taxonomy  when applied to a hierarchical topic. The second evaluation analyzes the topic observed coherence, considering that we are also  interested in the interpretability of the selected label sets along the hierarchy.

An important point is that there are many efficient and effective clustering solutions, so it could be good to have a cluster labeling solution independent from the clustering construction. Thus, we examined some labeling methods based on the ranking of the attributes in each cluster such as the MTWL$_{raw}$, a variation of it using ``idf'' as weight, and some using the ``JSD'' and the $\chi^2$ statistics. Also, we used the methods proposed by \citeN{Lamirel:2008}, which rank the attributes in a modified form of the \textit{F measure} calculation. These methods claim to be good for hypertree visualization, and they also claim to produce good specific and generic labels. Finally, we used some methods especially constructed to consider the hierarchical relationships among the topics, as proposed by \citeN{Muhr:2010}, \citeN{PopeUngar:2000} and \citeN{MouraRezende2010}. This last set of methods does not only rank the attributes, but makes a specific hierarchical attribute selection in order to create a topic hierarchy.

The results of the experimental comparison revealed that the labeling methods which consider the hierarchical relations presented good results in both evaluations. On the other hand, methods as simple as MTWL$_{raw}$, those based on JSD or $\chi^2$, also had good scores. Actually, the RLUM \cite{MouraRezende2010} and Popescul\&Ungar \cite{PopeUngar:2000} methods resulted in a stable behavior of the \textit{$F_{measure}$ values} curve in all retrieval tests, even when the dendrogram was unbalanced. Methods which use a hierarchical weight to select the label sets \cite{Muhr:2010} cause a smoothing of the \textit{$F_{measure}$ values} curve in comparison to their version without the hierarchical weights. However, these last results are better when the hierarchy is almost balanced. On the other hand, methods based on the RCL strategy, that is, considering only a local hierarchy, and those using JSD or $\chi^2$ to aid the label ranking, always present the best mean values of \textit{$F_{measure}$}. However, these methods are not stable in all types of hierarchies. The methods proposed by \cite{Lamirel:2008} had good results from the first to the middle levels of the hierarchy, as expected. These methods were proposed to improve the hypertree visualization of the clusters, so they had to be good for short hierarchies. 

In the coherence evaluation there were some different results. The methods which use a hierarchical weight \cite{Muhr:2010} presented the best values for the upper quartile, in comparison to the others. This can be evidence that the hierarchical weight brings attributes to the label set which carries some context information, as they claim. Thus, probably in a subjective evaluation, these methods could lead to the best classifications. The other methods were divided into two other groups. In the second group, there are methods based only on frequency or $\chi^2$, which seem to be efficient at obtaining the maximum coherence values. Finally, the last group encompasses more radical methods, such as Popescul\&Ungar e and RLUM, which produce very specific and very generic labels, without attribute replications. In the third group, methods use JSD or the $\chi^2$ estimation to make decisions. Thus, the simple frequencies combined with a hierarchical weight produced the most coherent topics - HierMTWL$_{idf}$, HierMTWL$_{raw}$, HierICWL$_{raw}$ and HierICWL$_{idf}$.  

Considering the interpretability and the exploratory browsing of the topic hierarchy, it would be great to have a balance between the best retrieval \textit{$F_{measure}$ values} and the observed coherence. The methods which use the hierarchical weights seemed to be closer to reaching this balance. However, they are not always the best choice. In the experiments there is evidence that, for unbalanced hierarchies, it is still better to use a robust method, such as RLUM or even the simplest one - MTWL$_{raw}$ (Most Frequent). 

The proposed comparison methodology can be extended to aid users in the task of selecting a useful topic hierarchy. The focus here is to cover the analysis of the label selection step, assuming well-defined scenarios of text preprocessing and clustering. Good topic hierarchies are a combination of well-chosen text representation, reliable hierarchical structure and a representative set of labels for each topic. In this way, some of the future works relies on different labeling methods, cluster algorithms and prune, as well as a different preprocessing or a more restrictive calculation for the coherence measure. These different processes does not affect the proposed comparison methodology, although they can obtain different results from the experiment presented in this paper. To actually enriched the methodology we can add a human evaluation. As discussed before, this human evaluation can be  guided by the simulation already done, in order to decrease the number of nodes to be criticized.

Furthermore, we provided a hierarchical cluster labeling benchmark with standardized algorithm specifications and the implementations\footnote{Details available on \url{http://sites.labic.icmc.usp.br/toptax/}.}. These standards should allow for aa fairer comparison between new methods and the existing ones. The experimental comparison can be extended by including other evaluation measures in order to compare the methods while considering different perspectives.

\begin{acknowledgements}
This research was supported by Embrapa Agricultural Informatics (project number: 01.12.01.001.03.00), the Brazilian Research Councils CAPES (process number: DS-6345378/D) and FAPESP (processes number: 2011/19850-9 and 2014/08996-0). Additionally, the authors would like to thank Anandsing Dwarkasing and Alexandre Romero Inforzato for the English revision of this paper. 
\end{acknowledgements}

\bibliographystyle{spbasic}      
\bibliography{MFMBIB}

\end{document}